Mono- and Polyauxic Growth Kinetics: A Semi-Mechanistic Framework for Complex Biological Dynamics

Gustavo Mockaitis[1]

[1]Interdisciplinary Research Group on Biotechnology Applied to the Agriculture and the Environment, School of Agricultural Engineering, University of Campinas (GBMA/FEAGRI/UNICAMP), 501 Cândido Rondon Avenue, CEP, 13.083-875, Campinas, SP, Brazil.

Corresponding author: gusmock@unicamp.br

ORCID 0000-0002-4231-1056

**Abstract**

Kinetic modeling of microbial growth is essential for the design, optimization, and scale-up of industrial bioprocesses. Classical empirical models often lack biologically interpretable parameters or fail to capture complex multiphasic (polyauxic) behaviors, while fully mechanistic models are impractical for systems involving complex substrates and mixed cultures. This study proposes a unified mathematical framework that reformulates the canonical Boltzmann and Gompertz equations into semi-mechanistic forms, explicitly defining the maximum specific reaction rate and lag phase duration. Polyauxic growth is represented as a weighted sum of sigmoidal phases, subject to stringent constraints that ensure parameter identifiability, temporal consistency, and biological plausibility. The methodology integrates a workflow to address nonlinear regression in high-dimensional parameter spaces. A two-stage optimization strategy using Differential Evolution for global search followed by L-BFGS-B for local refinement avoid bias and heuristic parameter initialization. A Charbonnier loss function and the Robust Regression and Outlier Removal procedure are employed to identify and mitigate experimental outliers. Model parsimony is enforced using Akaike (AIC, AICc) and Bayesian (BIC) information criteria to select the optimal number of growth phases and avoid overparameterization. The framework was evaluated using experimental anaerobic digestion datasets, demonstrating that conventional single-phase models can obscure relevant metabolic transitions in co-digestion systems.

## 1 Introduction

Analysis of microbial growth curves is fundamental in microbiology and biotechnology, supporting advances from basic research to industrial and environmental applications. Growth curves are typically measured as changes in biomass, substrate concentration, or product formation over time, and provide insight into microbial activity and bioprocess dynamics, although they serve only as indirect proxies for physiological or metabolic states (Armstrong 2008; Fiksen et al. 2013). Accurate modeling and interpretation of these curves are essential for bioreactor design, process optimization, and understanding ecological interactions (Pavlostathis and Giraldo-Gomez 1991). However, environmental and biotechnological systems frequently show microbial growth under conditions where substrate composition is complex or poorly characterized and microbial communities are inherently heterogeneous and dynamic. In such systems, classical assumptions of pure cultures growing on well-defined substrates are rarely satisfied, and growth dynamics often exhibit multi-phasic behavior arising from emergent population-level processes rather than identifiable substrate-specific pathways (Nielsen and Villadsen 1992; Kovárová-Kovar and Egli 1998).

Existing approaches to microbial growth modeling can be broadly categorized into (i) mechanistic and structured models, which explicitly represent intracellular regulation and substrate-specific pathways (Nielsen and Villadsen 1992; Ramkrishna and Song 2012; Bate et al. 2023), (ii) cybernetic models that assume optimal regulatory strategies (Kompala et al. 1986; Liu et al. 2000), and (iii) empirical or phenomenological models that describe observed growth patterns without resolving underlying mechanisms (Tsoularis and Wallace 2002; Escalante et al. 2017; Tjørve and Tjørve 2017). Although mechanistic and cybernetic models have demonstrated remarkable predictive capability under controlled conditions, their applicability relies on assumptions that limit their use in complex systems. These models require explicit knowledge of substrate identity, metabolic pathways, regulatory mechanisms, and often assume physiological homogeneity within the microbial population (Kompala et al. 1986). Such requirements are rarely met in systems involving complex substrates, mixed microbial consortia, or environmental samples, where kinetic parameters lose strict mechanistic identifiability (Kovárová-Kovar and Egli 1998) and become context-dependent (Fiksen et al. 2013).

In batch cultures, microbial growth generally follows a sigmoidal pattern, characterized by lag, exponential (log), stationary, and sometimes decline phases. The decline phase, although often omitted in engineering analyses, is biologically relevant in situations involving nutrient depletion, product inhibition, or cell death dynamics. The parameters commonly derived from the lag and exponential phases, such as the maximum specific growth rate, lag phase duration, and carrying capacity (asymptotic maximum level), are therefore not only useful for engineering design and process control, but also scientifically meaningful for describing the physiological behavior and adaptive responses of microbial populations. Empirical and non-mechanistic models such as the Boltzmann (Boltzmann 1872) and Gompertz (Gompertz 1825) equations are widely used as baseline frameworks for modeling microbial growth curves, due to their simplicity, robustness, and limited data requirements. Modifications to these equations, replacing mathematical coefficients with biologically interpretable parameters, established the standard for monoauxic growth analysis. These equations are commonly applied to fit growth data and extract such parameters (Zwietering et al. 1990).

From a mathematical standpoint, classical sigmoidal models such Boltzmann and Gompertz form a continuum rather than a set of unrelated equations. The Verhulst logistic, Gompertz, von Bertalanffy, Richards, Blumberg hyperlogistic and several other formulations can all be derived as special cases of a generalized logistic differential equation, in which shape exponents control the location of the inflection point, the symmetry of the curve and the decay of the relative growth rate. Within this broader family, the Boltzmann and Gompertz equations used here can be viewed as two limiting but practically important shapes: a symmetric transition with a narrow region of high rates and an asymmetric transition with an extended deceleration tail (Tsoularis and Wallace 2002).

Monoauxic growth curves, observed with single substrates, display a single sequence of growth phases (Jason 1983; Volpi et al. 2024). However, systems with mixed substrates, complex feedstocks, or mixed microbial communities often exhibit polyauxic (multiphasic) growth (Monod 1941, 1945; Blaiseau and Holmes 2021). Distinct sequential growth phases correspond to the preferential consumption of different substrates or substrate fractions, producing stacked sigmoidal curves. This phenomenon is especially relevant in environmental biotechnology, biorefineries, and processes involving metabolically versatile or mixed cultures (Zhang et al. 1998; Mockaitis et al. 2020; Champaneria et al. 2022).

Differentiating monoauxic from polyauxic growth is essential for accurately determining kinetic parameters and for the rational design and scale-up of bioreactors. In monoauxic systems, classical kinetic models allow reliable estimation of key parameters such as maximum specific growth rate, substrate affinity constant, yield coefficients, and maintenance requirements. However, in polyauxic systems, which are common in complex substrates or mixed cultures, each phase is governed by distinct substrate consumption kinetics and regulatory mechanisms. Applying a monoauxic model to such systems can yield oversimplified or misleading parameter estimates, compromising process optimization. Therefore, extraction of kinetic parameters from well-fitted growth curves is indispensable for the rational design, scale-up, and control of industrial bioprocesses, informing substrate loading rates, reactor sizing, and operational strategies. Reliable estimation of these parameters is a prerequisite for engineering efficient, reproducible, and economically viable processes.

While classic diauxic models describe defined transitions in pure cultures on simple substrates (e.g., glucose-lactose shifts), environmental and industrial bioprocesses often involve complex, undefined substrates (e.g., wastewater, biomass hydrolysates). In such systems, growth curves exhibit multi-phasic behaviors that are not strictly mechanistic switches but overlapping consumption kinetics. Traditional low-parameter models fail to capture these dynamics, necessitating robust semi-mechanistic frameworks capable of describing polyauxic profiles in complex matrices (Kompala et al. 1986; Liu et al. 2000). Identifying and modeling polyauxic growth allows for extraction of phase-specific kinetic parameters and deepens understanding of the physiological and ecological factors influencing microbial kinetics, substrate preference, and metabolic regulation. Recognizing and mathematically differentiating polyauxic from monoauxic growth is a significant step toward quantitative, predictive, and scalable engineering of microbial processes with complex feedstocks or consortia.

Reliable estimation of kinetic parameters relies on nonlinear regression analysis aimed at minimizing a suitable objective function, typically the residual sum of squares. While derivative-based algorithms like the classical Levenberg-Marquardt method are widely used, they are prone to entrapment in local minima, particularly in the high-dimensional and multimodal parameter spaces characteristic of polyauxic models. If the initial guess is too far from the global minimum, deterministical methods may fail to identify the true kinetic parameters. To overcome these limitations, global heuristic strategies have been successfully applied to ensure robust convergence to the global optimum, often serving as a robust initialization step for subsequent local refinement (Escalante et al. 2017).

Recent work has highlighted pervasive inconsistencies in how Gompertz parameters are interpreted. Many “modified Gompertz” formulations used in ecology and predictive microbiology misidentify the growth coefficient as a maximum relative growth rate, and common reparameterisations implicitly fix the lag time as a constant fraction of the asymptotic value rather than allowing it to emerge as an independent biological quantity (Tjørve and Tjørve 2017). Accordingly, maximum specific rates and lag phase duration should be derived directly from the analytical geometry of the fitted curve (particularly from the inflection point and its neighbourhood) thereby maintaining consistency with the kinetic interpretation of parameters in the Monod formalism. This distinction becomes important when growth models are used as quantitative tools for bioprocess design and control rather than as purely empirical curve fits.

Therefore, this work presents a methodological framework that redefines the application of the Boltzmann and Gompertz equations for microbial growth analysis. This approach systematically transforms these traditional empirical models into biologically meaningful tools, providing explicit physiological interpretation for each parameter. The method details how to generalize single sigmoid curves to n-auxic models, enabling consistent modeling of complex, multiphasic (polyauxic) growth behavior. The workflow also incorporates strategies for outlier elimination, objective curve fitting that avoids dependence on arbitrary initial guesses, and clear criteria to define the number of stacked sigmoids required to describe the microbial growth accurately, but preventing model overparameterization. These methodological advances ensure that the resulting kinetic models are not only statistically robust and reproducible, but also mechanistically informative and directly relevant to bioprocess design, optimization, and scale-up. The contribution of this work is the development of a unified semi-mechanistic framework for mono- and polyauxic growth, in which classical sigmoidal functions are reformulated to yield parameters with direct biological interpretation. This framework enables the description of multiphasic growth dynamics in systems where mechanistic modeling is impractical, while retaining interpretability beyond purely empirical fitting.

## 2 Theoretical Framework

The main challenge in bridging classical microbial kinetics with empirical models is to bring a consensus how traditionally defined kinetic parameters emerge from, and are constrained by, experimental observations. Most mathematical models for microbial kinetic rely on the Monod equation (Monod 1949), which defines apparent kinetic parameters, under fixed conditions, such as the maximum specific growth rate ($\mu_{max}$) and the half-saturation constant ($K_M$). These parameters are treated here as effective descriptors defined under fixed experimental conditions, acknowledging that in complex substrates or mixed cultures they do not retain strict mechanistic identifiability. These parameters are not explicitly time-dependent (within fixed experimental conditions) and are distinct from the maximum rate values obtained when fitting growth or substrate consumption curves at a given initial substrate concentration ($S_i$) (Lobry et al. 1992). While the Monod equation describes the biological relationship between growth rate and substrate concentration, the maximum rate measured in any single experiment is specific to the $S_i$ used and does not represent the asymptotic system maximum under saturating conditions.

Monod equation, assuming growth-associated processes, describes microbial growth, substrate consumption and products formation, and can be rewritten as shown in Equation 1 (Pirt 1982):

$$\mu(t) = \frac{Y_{X/P} \cdot r_P(t)}{X(t)} = \frac{Y_{X/S} \cdot r_S(t)}{X(t)} - m_S \qquad \text{Equation 1}$$

where $\mu(t)$ is the specific microbial growth rate at instant $t$; $r_P(t)$ is the instantaneous rate of product formation; $r_S(t)$ is the instantaneous rate of substrate consumption; $X(t)$ is the biomass concentration; $Y_{X/P}$ biomass yield coefficient on product; $Y_{X/S}$ biomass yield coefficient on substrate, and $m_S$ is the specific maintenance coefficient. At the point of maximum rate, Equation 1 can be written as Equation 2:

$$\mu_{max} = \frac{Y_{X/P} \cdot r_{P_{max}}}{X_f} = \frac{Y_{X/S} \cdot r_{S_{max}}}{X_f} - m_S \qquad \text{Equation 2}$$

where $r_{P_{max}}$ is the maximum rate of product formation; $r_{S_{max}}$ is the maximum rate of substrate consumption, and $X_f$ is the corresponding maximum biomass concentration.

The constants $r_{P_{max}}$, $r_{S_{max}}$ and $\mu_{max}$ can be generalized as $r^*_{max}$ in Monod equation, shown in Equation 3:

$$r(S) = r^*_{max} \cdot \frac{\mathrm{S}}{K_M + \mathrm{S}} \qquad \text{Equation 3}$$

where $r(S)$ is the instantaneous rate. However, $r^*_{max}$ represents the theoretical maximum rate, achieved only at saturating substrate concentrations. In a microbial growth experiment, the initial substrate concentration is often below the saturating value. Therefore, $r^*_{max}$ cannot be directly measured in a single batch experiment. Instead, the highest rate observed experimentally is determined by substituting $S = S_i$ into the Monod equation, resulting in the maximum apparent rate ($r^{app}_{max}$), as shown in Equation 4:

$$r^{app}_{max} = r_{max} \cdot \frac{S_i}{K_M + S_i} \qquad \text{Equation 4}$$

This shows that $r^{app}_{max}$ is not an intrinsic property, but rather a function of $S_i$. Moreover, considering very high inoculum concentrations or immobilized inoculum, apparent rate also includes internal and external mass transfer resistances limitations (Zaiat et al. 1997; Enouy et al. 2022), which are not addressed in this study.

To determine kinetic constants such as $r_{max}^*$ and $K_M$, multiple experiments must be conducted with varying $S_i$ values. For each condition, the maximum observed rate should be determined, to generate a dataset of $\left(r_{max}^{app}, S_i\right)$ pairs. Fitting these data to Equation 4, considering that $r_{max}^* = \lim_{S_i \to +\infty} r_{max}^{app}$, it is possible to determine the parameters $r_{max}^*$ and $K_M$. Integrating substrate consumption and growth profiles further enables kinetic parameters to be related to stoichiometric yields and maintenance coefficients, resulting in a comprehensive quantitative description of microbial systems. This approach provides a mathematical basis for extracting kinetic parameters and highlights the limitations of single-condition experiments, which can yield misleading or non-generalizable results. Microbial systems are often affected by substrate or product inhibition, maintenance requirements, and physiological state variations, all of which impact observed kinetics. Only systematic experimentation across a range of substrate concentrations can provide insight into the underlying kinetics of microbial growth and substrate utilization, disentangling microbial properties from process limitations. Therefore, a phenomenological-mechanistic continuum framework, where empirical descriptions are progressively enriched with physical and biological interpretation is a possible solution for complex systems kinetics without requiring full mechanistic resolution.

## 3 Monoauxic Modified Boltzmann Equation

The Boltzmann function, a classic sigmoidal equation, was proposed by Ludwig Boltzmann (Boltzmann 1872). It can be viewed as a modification of the logistic function and is presented in Equation 5. This model mathematically describes systems undergoing transitions between two discrete states, such as in thermodynamics and phase transitions. The Boltzmann equation yields a symmetric sigmoidal curve centered at its inflection point, such that the rates of change before and after this inflection point are identical. This symmetry makes it especially suitable for modeling physical or chemical processes where the transition between initial and final states occurs at similar rates in both directions. Equation 5 shows the canonical form of Boltzmann equation:

$$y(x) = y_i + \frac{\left(y_f - y_i\right)}{1 + e^{\left(\frac{x_0 - x}{\gamma}\right)}} \qquad \text{Equation 5}$$

where $y_i$ is the asymptotic minimum value reached by the ordinate; $y_f$ is the asymptotic maximum value reached by the ordinate; $x_0$ is the abscissa at which the slope attains its maximum, defined by $x_0 = x \Leftrightarrow y(x_0) = \frac{(y_f + y_i)}{2}$; and $\gamma$ is a parameter correlated with the slope at the inflection point. The term $\left(y_f - y_i\right)$ is the amplitude factor of the function, in which $y_f = \lim_{x \to +\infty} y(x)$ and $y_i = \lim_{x \to -\infty} y(x)$.

Although $\gamma$ is associated with the steepness of the transition, it additionally possesses a clear physical interpretation, representing a characteristic time scale that marks the 27% and 73% amplitude points around the inflection time. To provide a more directly meaningful parameter, the first derivative of the Boltzmann equation is considered, representing the rate of change at each point (Equation 6):

$$r(x) = \frac{dy(x)}{dx} = \frac{(y_f - y_i)}{\gamma} \cdot \frac{e^{\left(\frac{x_0 - x}{\gamma}\right)}}{\left[1 + e^{\left(\frac{x_0 - x}{\gamma}\right)}\right]^2} \qquad \text{Equation 6}$$

Since the first derivative of any sigmoidal function is a peak-shaped function, its maximum value ($r_{max}$) occurs at the point where its second derivative (Equation 7) equals zero, as illustrated in Figure 1. $r_{max}$ quantifies the maximum rate of change during the transition between states and is meaningful in both physical and biological systems. In phase transitions or chemical reactions, $r_{max}$ marks the point of greatest activity. In biological systems, it reflects the highest rates of growth, response, or product formation, serving as a key indicator of system dynamics and efficiency.

$$\frac{d^2y(x)}{dx^2} = \frac{(y_f - y_i)}{\gamma^2} \cdot \frac{e^{\left(\frac{x_0 - x}{\gamma}\right)} \cdot \left[e^{\left(\frac{x_0 - x}{\gamma}\right)} - 1\right]}{\left[1 + e^{\left(\frac{x_0 - x}{\gamma}\right)}\right]^3} \qquad \text{Equation 7}$$

Figure 1 illustrates the behavior of a generic Boltzmann sigmoid and its first and second derivatives (Equation 5 to Equation 7, respectively):

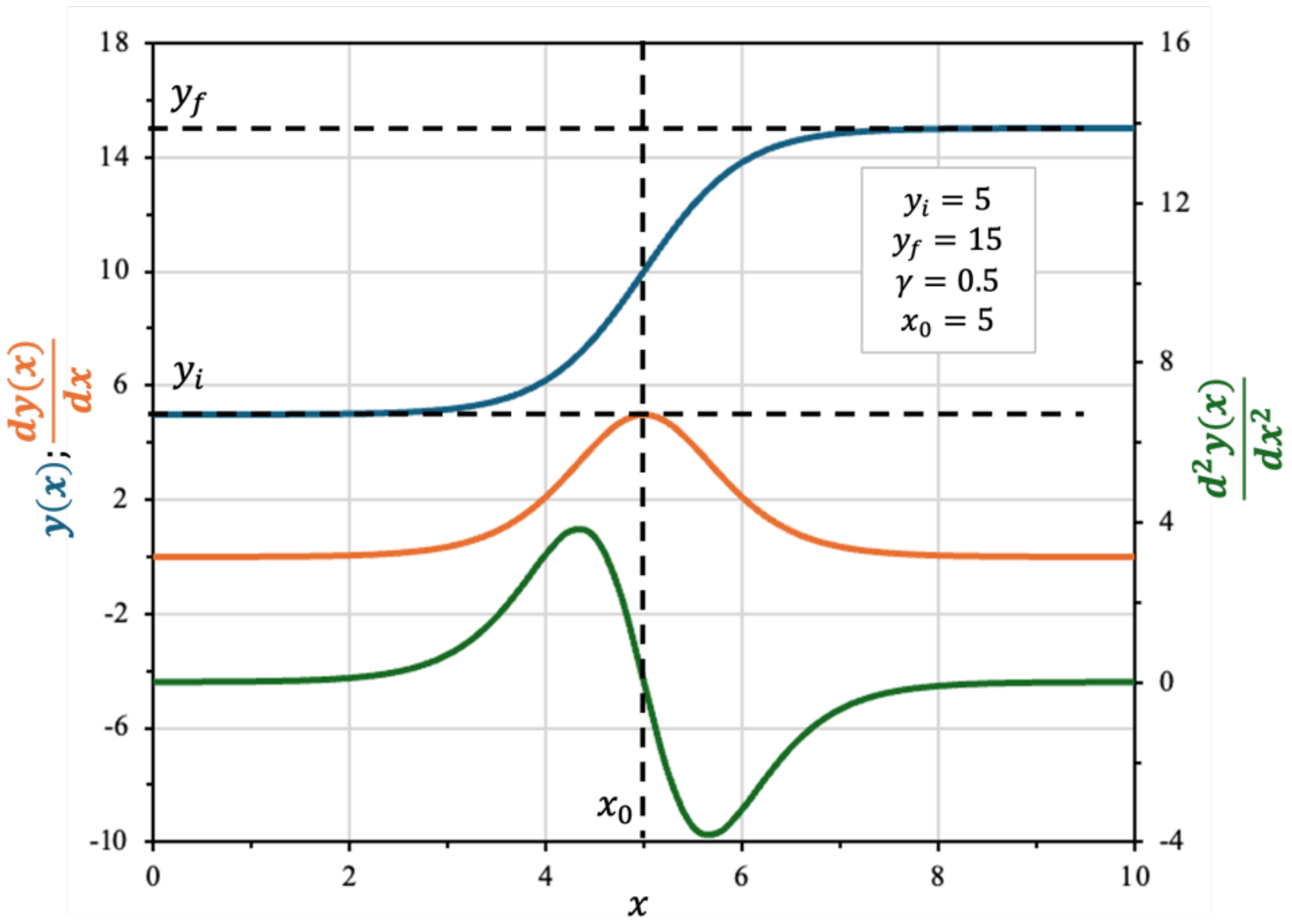

Figure 1 – Boltzmann sigmoid (plotted in blue) and its first (orange) and second (green) derivatives.

The point where the second derivative is zero (Equation 7) is the maximum point of its first derivative (Equation 6). This point corresponds to the maximum slope ($r_{max}$). Equation 7 is a transcendental equation which is not reducible to a polynomial canonical form, due to the impossibility to isolate the independent variable. Hence, the only possible value that satisfies the solving value of $x$ for Equation 7 root is $x_0$ for $y_i, y_f, \gamma \neq 0$. This confirms that $x_0$ corresponds to the abscissa at which the slope of Equation 5 is maximum. Thus, replacing $x$ by $x_0$ in Equation 6 leads to Equation 8, which express $r_{max}$ as a function of $\gamma$:

$$r_{max} = r(x_o) = \frac{(y_f - y_i)}{4 \cdot \gamma}$$ Equation 8

Isolating $\gamma$ in Equation 8 and replacing it in Equation 5, leads to Equation 9 which incorporates $r_{max}$ as a parameter:

$$y(x) = y_i + \frac{(y_f - y_i)}{1 + e^{\left(\frac{4 \cdot r_{max} \cdot (x_0 - x)}{(y_f - y_i)}\right)}}$$ Equation 9

This model does not have a parameter associated with the time length of the lag phase ($\lambda$). $\lambda$ can be calculated assuming that the exponential growth phase (log phase) has a quasi-linear behavior where $r_{max}$ is its angular coefficient, thus:

$$y^*(x) = r_{max} \cdot (\lambda - x_0) + y(x_0)$$ Equation 10

The time length of the lag phase corresponds to the time when $y^*(x) = y_i$:

$$y_i = r_{max} \cdot (\lambda - x_0) + y_i + \frac{y_f - y_i}{2} \Rightarrow x_0 = \lambda + \frac{y_f - y_i}{2 \cdot r_{max}}$$ Equation 11

Replacing $x_0$ defined in Equation 11 in Equation 9 leads to a general equation depicted by Equation 12:

$$y(x) = y_i + \frac{(y_f - y_i)}{1 + e^{\left(\frac{4 \cdot r_{max} \cdot (\lambda - x)}{(y_f - y_i)} + 2\right)}}$$ Equation 12

This reparameterized expression is mathematically equivalent to the modified Logistic model, which was established as a standard form to facilitate biological interpretation (Zwietering et al. 1990). To illustrate the reparameterization of Equation 5, rewrote in terms of $r_{max}$ and $\lambda$, and correlates it with the quasi-linear approximation of the exponential growth region, Figure 2 shows the behavior of both Equation 11 (the linearized tangent at the inflection point) and Equation 12 (the full sigmoidal model). This comparison highlights how the tangent line approximates the sigmoid's exponential region and defines the lag phase duration.

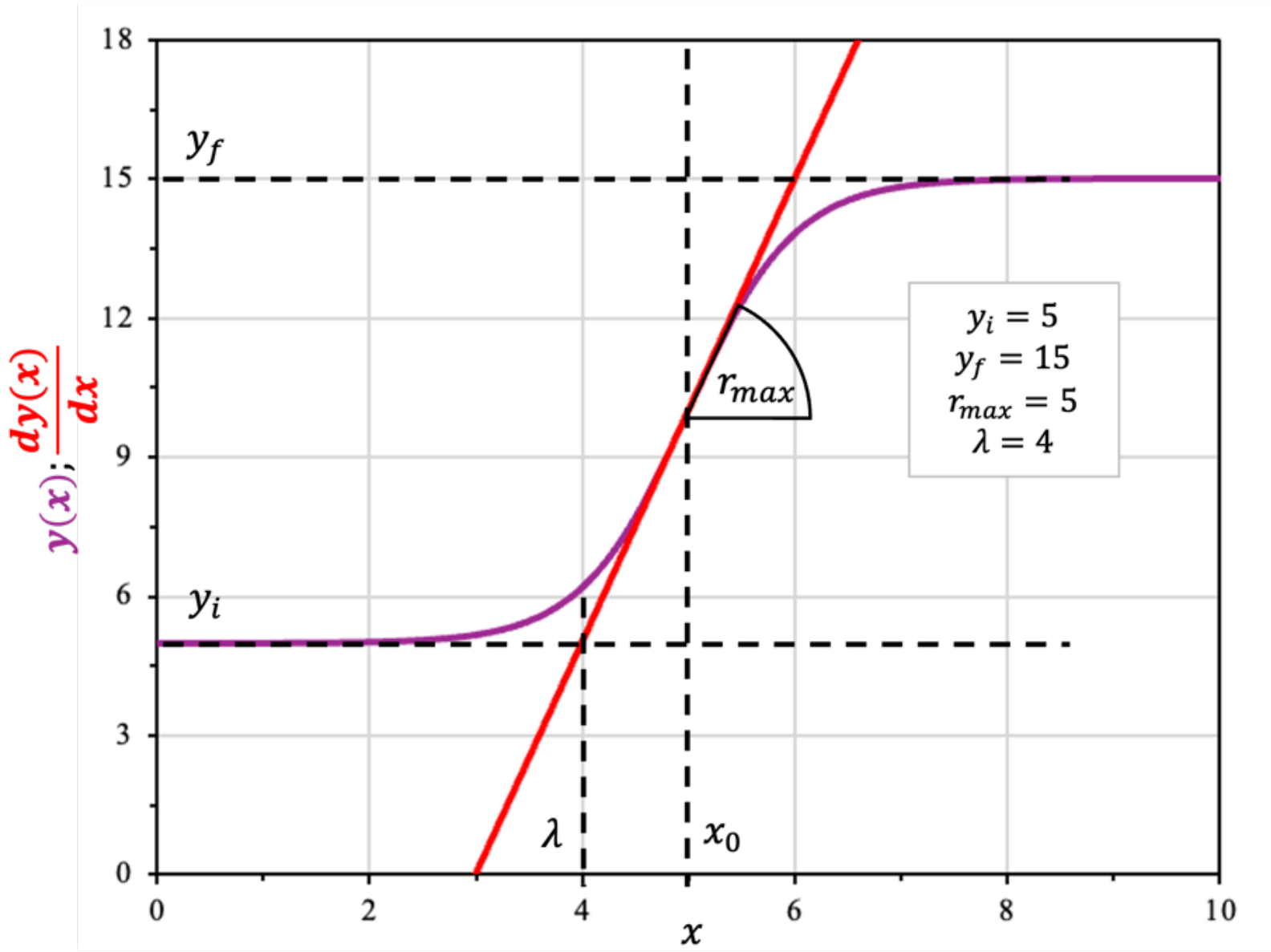

Figure 2 – Modified Boltzmann sigmoidal model (Equation 12; plotted in purple) and the corresponding tangent at the inflection point as given by (Equation 11; in red).

Variables and parameters in Equation 12 can be reassigned to reflect biological significance; therefore the independent variable $x$ is identified with time $(t)$. Equation 13 shows the model expressed as parameters of production of a product formation $P$. Since, in experimental assays, the initial product concentration is zero, this simplifying hypothesis can be applied:

$$P(t) = \frac{P_{max}}{1 + e^{\left(\frac{4 \cdot r^{app}_{P_{max}} \cdot (\lambda - t)}{P_{max}} + 2\right)}}$$ Equation 13

Since substrate is seldom depleted in a biological process, it also possible to express the behavior of substrate concentration, using Equation 8, which can be rewritten into Equation 14:

$$S(t) = S_i + \frac{(S_f - S_i)}{1 + e^{\left(\frac{4 \cdot r_{S_{max}}^{app} \cdot (\lambda - t)}{(S_f - S_i)} + 2\right)}} \qquad \text{Equation 14}$$

As for microbial growth, it might be interesting express $r_{max}$ as maximum specific growth rate ($\mu_{max}^{app}$), as defined in Monod equation (Equation 4), which describes microbial growth in terms of specific growth rate ($\mu$), as shown in Equation 15:

$$\mu = \frac{1}{X(t)} \cdot \frac{dX}{dt} \qquad \text{Equation 15}$$

$\mu_{max}^{app}$ can be defined as $\mu$ when the derivative term of Equation 12 is maximum:

$$\mu_{max}^{app} = \frac{1}{X(t^*)} \cdot \frac{dX}{dt}\bigg|_{t^*} \Rightarrow r_{max} = \mu_{max}^{app} \cdot \frac{X_f - X_i}{2} \qquad \text{Equation 16}$$

where $t^*$ denotes the inflection time, corresponding to the quantity previously indicated as $x_0$ in the canonical Boltzmann form (Equation 5). Replacing Equation 16 in Equation 12 and reassigning all parameters to describe specific microbial growth:

$$X(t) = X_i + \frac{(X_f - X_i)}{1 + e^{(2 \cdot \mu_{max}^{app} \cdot (\lambda - t) + 2)}} \qquad \text{Equation 17}$$

These reparametrized forms of the Boltzmann function enhance the interpretability and applicability of the model for physical, chemical, and biological processes by aligning the parameters with experimentally relevant quantities.

## 4 Monoauxic Modified Gompertz Equation

The Gompertz function, first formulated by Benjamin Gompertz in 1825 (Gompertz 1825), was originally developed to model the exponential increase in human mortality with age. Unlike symmetric sigmoidal models (as Boltzmann equation), the Gompertz equation generates an asymmetric S-shaped curve featuring a steep initial phase followed by a prolonged, gradually slowing tail, which reflects processes where the rate of change decelerates exponentially over time. Because of this asymmetry, the Gompertz function is especially effective for describing biological phenomena such as microbial growth, tumor expansion, and demographic trends, where early rapid changes are followed by slower, sustained progression.

Over time, the Gompertz model has been widely applied in various fields to represent systems where the progression is not uniform across all phases. The key distinction from symmetric models (such as the Boltzmann or logistic functions) is that the Gompertz curve's inflection point does not divide the curve into two mirror-image halves; instead, the growth or decline phase is sharper on one side and more gradual on the other. This makes the Gompertz equation particularly suitable for empirical data sets where processes accelerate rapidly but decelerate slowly. The mathematical canonical form of the Gompertz function is presented in Equation 18.

$$y(x) = y_i + (y_f - y_i) \cdot e^{-b \cdot e^{-c \cdot x}} \qquad \text{Equation 18}$$

where $y_i$ is the asymptotic minimum value approached by the function; $y_f$ is the asymptotic maximum value approached by the function; $b$ determines the horizontal position of the curve and is associated with the timing of the inflection point, and $c$ controls the steepness, being directly related with the slope at the inflection point. Collectively, these parameters define both the position and the rate at which the sigmoidal transition occurs. The amplitude factor $(y_f - y_i)$ used in Gompertz function is the same as described for Boltzmann equation, depicted in Equation 1.

However, unlike the Boltzmann equation, in which only the parameter γ lacks direct physical or biological interpretation, in the canonical form of the Gompertz equation none of the parameters ($b$ or $c$) inherently possess clear physical or biological significance. For the Gompertz model to be useful in microbial growth kinetics, the original parameters $b$ and $c$ must be redefined in terms of biologically meaningful parameters, specifically as λ (lag phase time length) and $r_{max}$ (maximum rate), respectively. This reparameterization enables the Gompertz model to represent microbial growth dynamics in a way that is directly interpretable in biological terms. To reassign $c$ as $r_{max}$, the rate function is defined through the first derivative of Equation 18:

$$r(x) = \frac{dy(x)}{dx} = (y_f - y_i) \cdot b \cdot c \cdot e^{-c \cdot x} \cdot e^{-b \cdot e^{-c \cdot x}}$$ Equation 19

The abscissa value in which $r_{max}$ occurs can be calculated as the root of the second derivative function (Equation 20).

$$\frac{d^2y(x)}{dx^2} = (y_f - y_i) \cdot b \cdot c^2 \cdot \left(b \cdot e^{b \cdot (-e^{-c \cdot x}) - 2 \cdot c \cdot x} - e^{b \cdot (-e^{-c \cdot x}) - c \cdot x}\right)$$ Equation 20

Figure 3 illustrates the behavior of a typical Gompertz curve alongside its first and second derivatives, corresponding to Equation 18 to Equation 20.

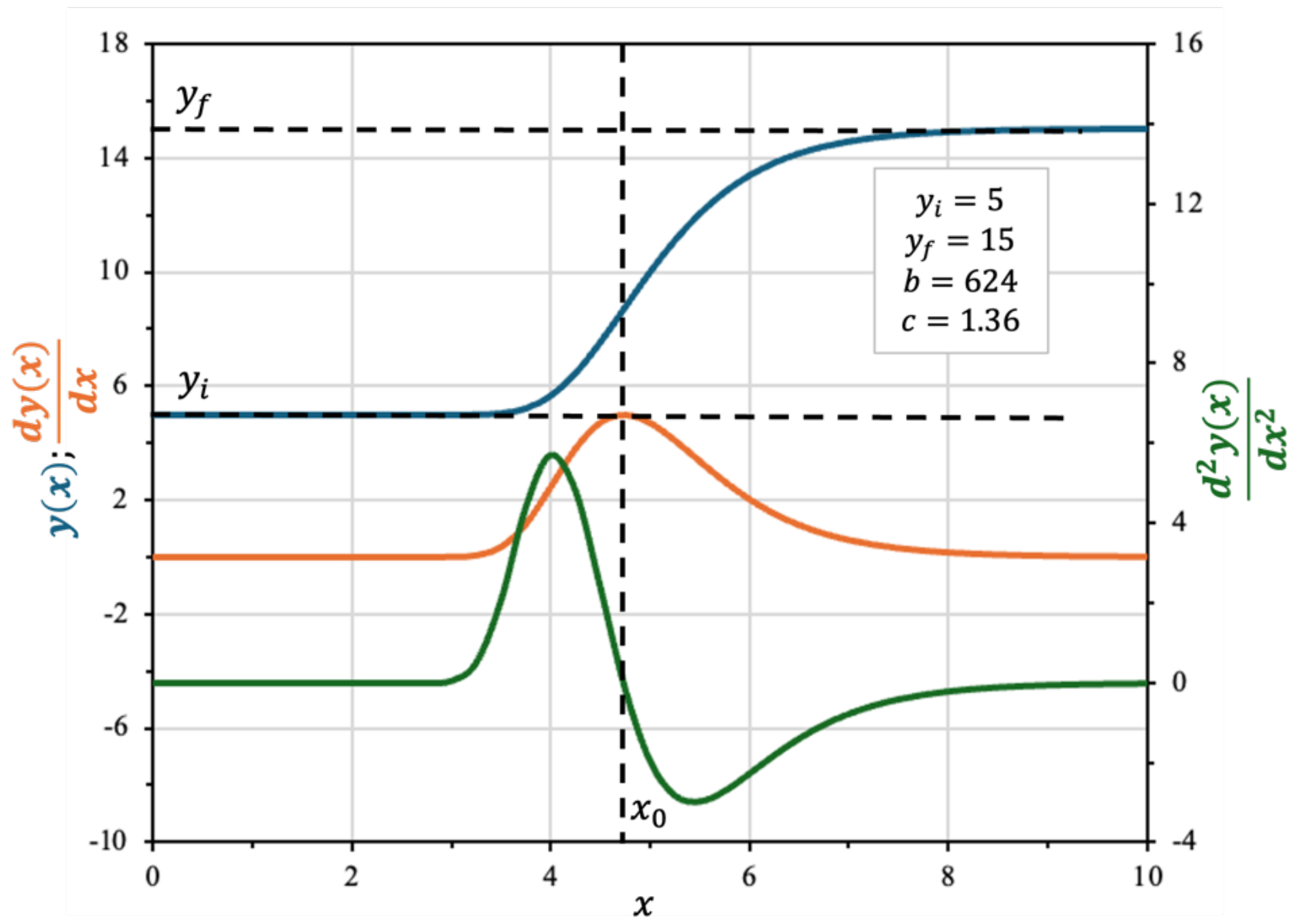

Figure 3 – Gompertz equation (plotted in blue) with its first derivative (orange) and second derivative (green).

The maximum of the first derivative, which represents the highest rate of change ($r_{max}$), occurs at the point where the second derivative equals zero. Unlike symmetric sigmoidal models such as the Boltzmann function, the inflection point of the Gompertz curve is not equidistant from the asymptotes, reflecting its characteristic asymmetry. In contrast to the Boltzmann equation, the second derivative of the Gompertz function can be algebraically simplified, allowing the root (inflection point) to be explicitly written as a function of the parameters $b$ and $c$. By solving the root and isolating $x$ value, hereafter reassign as $x_0$, Equation 21 is obtained:

$$x_o = \frac{\ln b}{c} \quad \text{Equation 21}$$

Replacing Equation 21 in Equation 19, is possible to define $r_{max}$:

$$r_{max} = r(x_o) = (y_f - y_i) \cdot \mathrm{c} \cdot e^{-1} \quad \text{Equation 22}$$

Isolating $c$ in Equation 21 and replacing it in Equation 18, leads to Equation 23 which incorporates $r_{max}$ as a parameter:

$$y(x) = y_i + (y_f - y_i) \cdot e^{-b \cdot e^{-\frac{r_{max} \cdot e}{(y_f - y_i)} \cdot x}} \quad \text{Equation 23}$$

As with the Boltzmann equation (Equation 9), the canonical form of the Gompertz model does not explicitly contain a parameter for the lag phase duration ($\lambda$). To redefine the parameter $b$ in terms of $\lambda$, a similar approach is adopted: the exponential growth (log) phase is approximated as a quasi-linear segment, with $r_{max}$ representing its slope, as described in Equation 10. This allows the lag phase duration to be directly incorporated into the Gompertz model, making its parameters biologically interpretable.

Rewriting the Equation 10 assuming that $\lambda$ corresponds to the time i.e. $y^*(\lambda) = y_i$, leads to Equation 24:

$$y_i = r_{max} \cdot \left(\lambda - \frac{\ln b}{c}\right) + y_i + \left(\frac{y_f - y_i}{e}\right) \Rightarrow b = e^{\left(1 + \lambda \cdot \frac{r_{max} \cdot e}{(y_f - y_i)}\right)} \quad \text{Equation 24}$$

Replacing the Equation 24 in Equation 22, leads to a reparametrized Gompertz equation, with $r_{max}$ and $\lambda$ as parameters, defined in Equation 25:

$$y(x) = y_i + (y_f - y_i) \cdot e^{-e^{\left(1 + \frac{r_{max} \cdot e}{(y_f - y_i)} \cdot (\lambda - x)\right)}} \quad \text{Equation 25}$$

This equation matches the modified Gompertz model, confirming that the geometric derivation via the inflection point yields the same robust kinetic tool widely used in predictive microbiology (Zwietering et al. 1990). Figure 4 displays the reparametrized Gompertz model alongside its linear tangent at the inflection point. This comparison demonstrates how the tangent approximates the rapid growth phase and clarifies the geometric interpretation of the lag phase parameter.

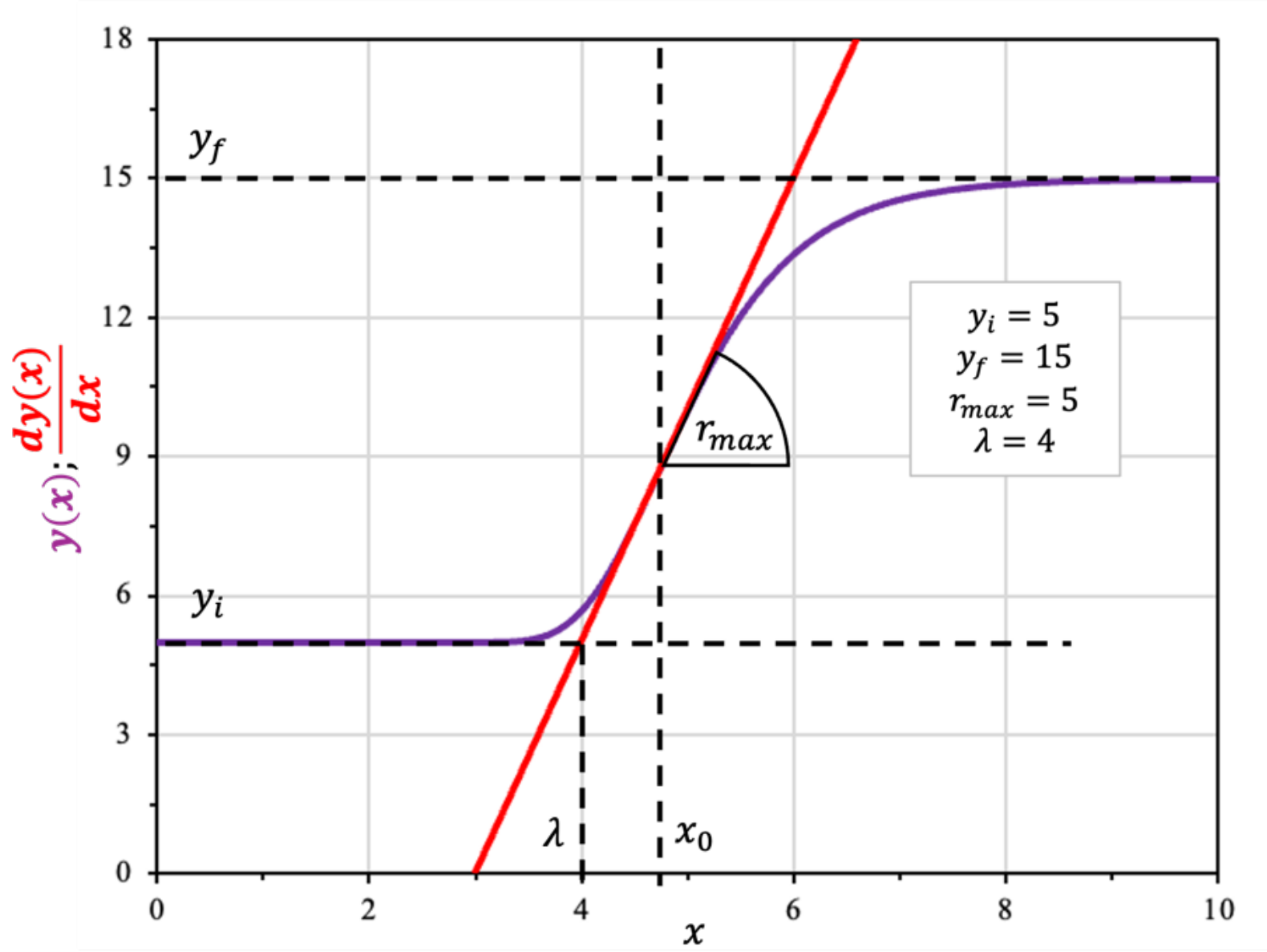

Figure 4 – Modified Gompertz sigmoidal model (Equation 25; plotted in purple) and the corresponding tangent at the inflection point as given by (Equation 24; in red).

To express the Gompertz function using biologically meaningful parameters, such as the maximum product concentration ($P_{max}$), the maximum product formation rate ($r_{P_{max}}^{app}$), and the lag phase duration ($\lambda$), the variables and parameters of the Gompertz equation are reassigned to reflect biological significance. Accordingly, the independent variable $x$ is identified with time ($t$) when the model is applied to time-resolved biological data. Under this formulation, the equation for product formation can be rewritten as:

$$P(t) = P_{max} \cdot e^{-e^{\left(1+\frac{r_{P_{max}}^{app} \cdot e}{P_{max}} \cdot (\lambda - t)\right)}} \qquad \text{Equation 26}$$

In a similar way, if substrate concentration is being tracked rather than product, the Gompertz model can be adapted as:

$$S(t) = S_i + (S_f - S_i) \cdot e^{-e^{\left(1+\frac{r_{S_{max}}^{app} \cdot e}{(S_f - S_i)} \cdot (\lambda - t)\right)}} \qquad \text{Equation 27}$$

For microbial growth, it is often useful to relate the maximum rate parameter ($r_{max}$) to the maximum specific growth rate ($\mu_{max}^{app}$). This can be accomplished by evaluating the specific growth rate at the inflection time $t^*$, which corresponds to the parameter $x_0$ in the canonical Gompertz representation:

$$\mu_{max}^{app} = \frac{1}{X(t^*)} \cdot \frac{dX}{dt}\bigg|_{t^*} \Rightarrow r_{max} = \mu_{max}^{app} \cdot \left(X_i + \left(X_f - X_i\right) \cdot e^{-1}\right) \qquad \text{Equation 28}$$

Substituting this relationship into the Gompertz function gives the reparametrized model for biomass concentration:

$$X(t) = X_i + \left(X_f - X_i\right) \cdot e^{-e^{\left(1+\mu_{max}^{app}\cdot\left(\frac{X_i \cdot e}{\left(X_f - X_i\right)}+1\right)\cdot(\lambda - t)\right)}} \qquad \text{Equation 29}$$

These reparametrized forms of the Gompertz equation directly express the model in terms of biologically relevant quantities such as lag phase, maximum specific growth rate, and asymptotic capacity, thereby enhancing interpretability in applied contexts. This idea is consistent with the reformulated Gompertz curve so that its parameters correspond to biologically meaningful descriptors (Zwietering et al. 1990). The present formulation extends this rationale by adopting a semi-mechanistic scaling strategy that explicitly accounts for nonzero initial conditions and defines the maximum specific growth rate based on the inflection value of the state variable, following the type of consistency considerations recently revisited (Guo and Wang 2024). As the formulation of modified Gompertz equation provides a biologically interpretable parameterization for single-phase growth (Zwietering et al. 1990), the present reformulation is particularly advantageous in polyauxic systems, because the amplitude $\left(X_f - X_i\right)$ remains explicitly associated with each growth phase. Therefore, this local amplitude scaling avoids embedding phase-dependent growth dynamics into global normalization constants, thereby improving interpretability and facilitating extension to n-auxic growth models. Moreover, this rationale also addresses inconsistencies highlighted in recent analyses of Gompertz reparameterisations, ensuring that lag and maximum rate derive objectively from curve geometry rather than being imposed through fixed empirical relationships (Tjørve and Tjørve 2017). This clarification is essential when comparing kinetic estimates across methodologies or relating sigmoidal growth descriptions to mechanistic Monod-type models in microbiology and biotechnology.

## 5 Polyauxic models

Polyauxic growth refers to the phenomenon where microorganisms exhibit multiple, sequential phases of growth when cultured in the presence of complex or multiple substrates. Unlike simple (monophasic) growth curves, polyauxic growth is characterized by the appearance of distinct growth phases, typically associated with the preferential consumption of a specific substrate or the metabolic adaptation to new resources as conditions change. This concept is particularly relevant for describing the growth dynamics of microbial cultures exposed to mixtures of substrates, such as in the degradation of complex organic matter or bioprocesses involving lignocellulosic feedstocks. Accurate modeling of polyauxic growth is important for understanding microbial physiology, optimizing industrial fermentations, and predicting substrate utilization patterns in environmental and biotechnological applications.

To describe microbiological polyauxic behavior, it is possible to model the overall growth curve as a weighted sum of multiple sigmoidal (or other suitable) functions. In this approach, the composite model is constructed as a summation over $j$, where each term in the sum corresponds to a distinct growth phase or substrate utilization event, represented by its own function indexed by $j$. Each of these component functions varies between its own initial and final ordinate values and is assigned a weighting factor $p_j$, which serves as a multiplicative coefficient scaling the contribution of that phase relative to the total amplitude of the data to be modeled. By ensuring that the sum of the weighting factors equals one, the model preserves the correct total amplitude and captures the sequential or overlapping phases characteristic of polyauxic growth.

To ensure that each component function indexed by $j$ contributes only a fraction of the total amplitude to the composite model, each amplitude term $\left(y_f - y_i\right)$ for each individual function should be multiplied by its respective weighting factor $p_j$. This means that, for each phase $j$, the weighted term $\left(y_f - y_i\right) \cdot p_j$ represents the effective amplitude contributed by that phase to the overall curve, as depicted in Equation 30:

$$y(x) = y_i + \left(y_f - y_i\right) \sum_{j=1}^{n} p_j \cdot f_j(x, p_j) \qquad \text{Equation 30}$$

Equation 30 describes the generic form of any stacked function describing polyauxic behavior, from any monoauxic function $f_j\left(x; \left(y_f - y_i\right)\right)$. Where $y_i$ and $y_f$ denote, respectively, the global initial and final values of the dependent variable, while each component function's amplitude $(y_f - y_i)$ is modulated by its weighting factor $p_j$.

If each phase is modeled by a Boltzmann-type sigmoidal function (as in Equation 12), the composite polyauxic curve becomes:

$$y(x) = y_i + (y_f - y_i) \cdot \sum_{j=1}^{n} \frac{p_j}{1 + e^{\frac{4 \cdot r_{max_j} \cdot (\lambda_j - x)}{(y_f - y_i) \cdot p_j} + 2}} \qquad \text{Equation 31}$$

Alternatively, if each growth phase is represented by a Gompertz-type function, the composite polyauxic model is expressed as:

$$y(x) = y_i + (y_f - y_i) \cdot \sum_{j=1}^{n} p_j \cdot e^{-e^{\left(1 + \frac{r_{max_j} \cdot e}{(y_f - y_i) \cdot p_j} \cdot (\lambda_j - x)\right)}} \qquad \text{Equation 32}$$

Figure 5 below illustrates an example of a three-phase polyauxic growth model, plotted using both modified Boltzmann (blue) and modified Gompertz (orange) sigmoidal equations. The key parameters for each phase are annotated on the graph. The dashed lines highlight the amplitude contributed by each phase, as determined by the weighted sum formalism. This example demonstrates how the proposed model structure can accurately describe complex, multi-phasic microbial growth behaviors and provides a visual reference for interpreting the meaning and roles of the fitted parameters.

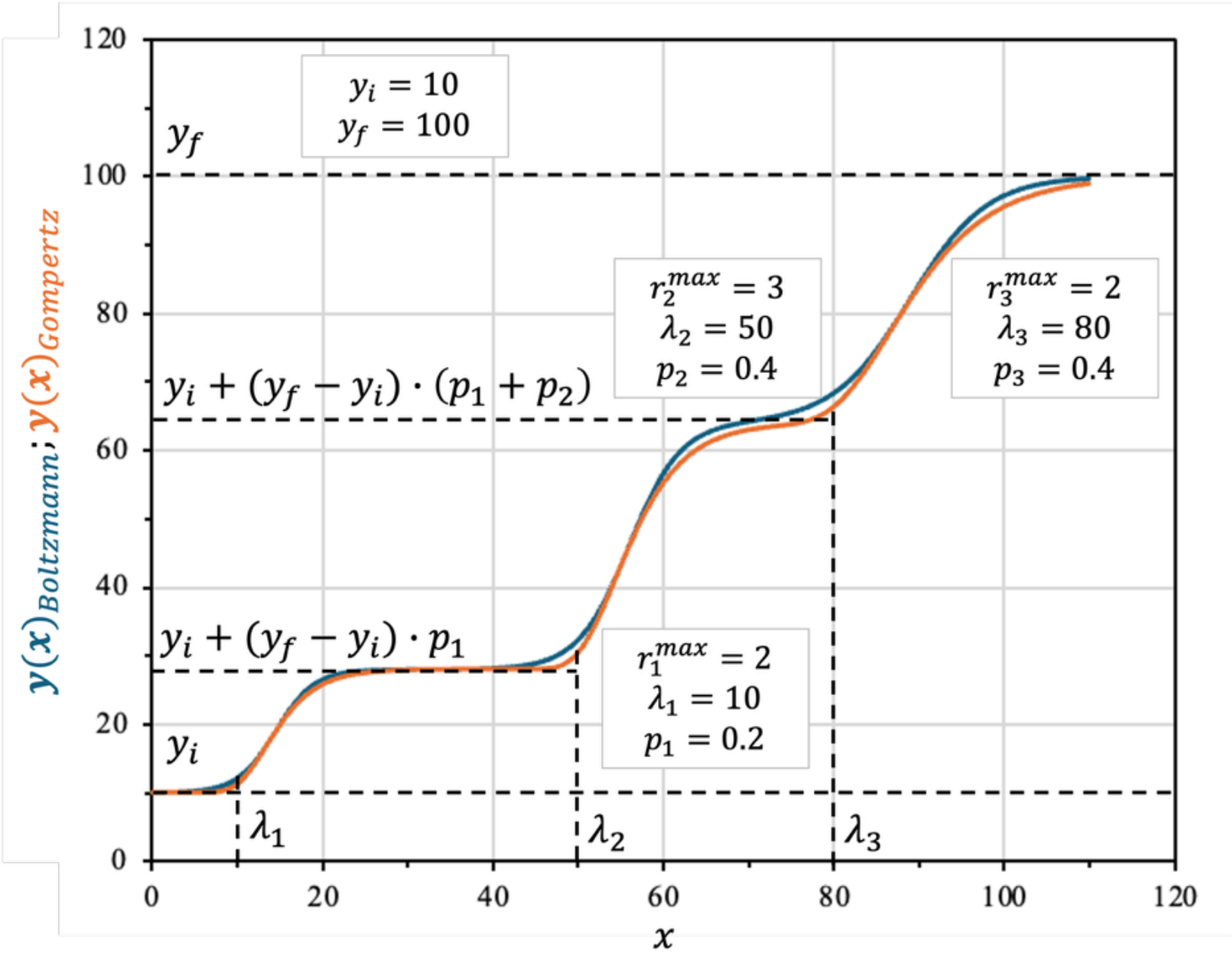

Figure 5 – Polyauxic growth for $n = 3$, using the modified sigmoidal kinetics model for Boltzmann (Equation 12; plotted in blue) and for Gompertz (Equation 25; in orange) equations.

To ensure all weighting factors are non-negative and sum to one (preserving interpretability and proper scaling), the weights can be parameterized using a softmax function:

$$p_j = \frac{e^{z_j}}{\sum_{l=1}^{n} e^{z_l}} \qquad \text{Equation 33}$$

where $z_j$ and $z_l$ are unconstrained parameters. The softmax transformation guarantees $0 < p_j \leq 1$ and $\sum_{l=1}^{n} p_l = 1$, ensuring that all weighting factors are non-negative and sum to 1. In addition to the softmax transformation constraining $p_j$, further conditions are defined to maintain model identifiability and to ensure the physical and biological plausibility of the fitted polyauxic curves. All maximum specific rates ($r_j^{max}$) are constrained to be strictly positive $r_j^{max} > 0$, reflecting the requirement that each growth phase or process exhibits a real, nonzero maximum rate. The first lag phase length ($\lambda_1$) must be zero or strictly positive ($\lambda_1 \geq 0$) and all subsequent lag phase parameter must be ordered such that $\lambda_{j+1} > \lambda_j$. This constraint enforces both chronological coherence among sequential phases and prevents degenerate solutions where multiple phases unrealistically overlap. Collectively, these constraints ensure that each component function contributes a well-defined, interpretable fraction to the overall amplitude and that the resulting composite model reflects both the mathematical structure and the physical realities underlying polyauxic growth behavior.

## 6 Model Fitting and Parameter Estimation

The estimation of model parameters for monoauxic and polyauxic growth curves requires a systematic, multi-stage approach to minimize bias from local minima and ensure robust convergence to the global optimum. Fitting sigmoidal models such as the modified Boltzmann and Gompertz equations, particularly in their polyauxic forms, necessitates non-linear regression algorithms capable of efficiently navigating complex, high-dimensional parameter spaces. As the number of phases ($n$) increases, especially when sigmoidal patterns overlap, the optimization landscape becomes increasingly complex, with heightened risks of parameter non-identifiability and convergence to suboptimal or meaningless solutions.

### 6.1 Fitting procedure

Overlapping sigmoids present a significant challenge, as their parameters can become highly correlated, making the fitting process sensitive to initial conditions. While constraints such as the non-negativity of rates and chronological ordering of phases help enforce model identifiability, they do not fully eliminate the risk of poor convergence to local minima. Achieving consistent and reproducible parameter values for a given dataset, using the same fitting method, is essential for scientific rigor. However, when parameter estimation relies on user-defined initial values, results may diverge substantially between runs as model complexity increases, often yielding trivial or non-interpretable solutions. Therefore, robust global search procedures and reduced reliance on arbitrary initial parameter values are crucial for reliable and interpretable fitting of polyauxic sigmoidal models.

To ensure numerical stability and minimize bias from scaling differences, the experimental data should be normalized by their maximum values. Subsequently, an automated heuristic initialization is performed, where the first derivative of the response data is analyzed to identify peaks corresponding to inflection points. These estimates provide a starting population for the optimization workflow, ensuring the search begins within a biologically plausible region of the parameter space.

To identify the optimal parameter set, Differential Evolution (DE) is employed as a robust, population-based global optimization algorithm (Storn and Price 1997). DE evolves a population of candidate solutions, utilizing mutation, crossover, and selection operators to explore the parameter space. This stochastic approach allows the algorithm to efficiently search for the global minimum of the loss function without being trapped by local optima or limited by specific user-defined starting values. This approach is particularly effective for high-dimensional, multi-modal problems such as fitting polyauxic sigmoidal models. In this sense, DE increases the likelihood of identifying parameter sets that are both optimal and physically or biologically meaningful, even in the presence of complex interactions and overlapping sigmoidal phases. As a result, DE provides a practical and reproducible approach to parameter estimation, ensuring the robustness of the fitted results.

After identifying a global solution with DE, the Limited-memory Broyden-Fletcher-Goldfarb-Shanno with Bounds algorithm (L-BFGS-B) is applied for local refinement (Byrd et al. 1995). L-BFGS-B is a gradient-based optimization method that efficiently converges toward a precise local minimum while strictly enforcing bound constraints on the parameters. This ensures that kinetic constants, such as maximum specific rates, remain strictly non-negative. Combining DE for global exploration with L-BFGS-B for local optimization ensures that parameter estimates are both robust and precise.

The application of the Lay-modified Gompertz model to biohydrogen production was assessed recently (Guo and Wang 2024). The authors provided step-by-step tutorials for fitting this model in several different commercial applications and their analysis confirms both the widespread adoption and the good descriptive performance of the single-phase Gompertz model in this field but also highlights that convergence and parameter estimates are highly sensitive to the initial values specified by the user. In their workflows, inadequate starting guesses often require manual trial-and-error adjustment before acceptable fits are obtained. By adopting a population-based global search (DE) followed by bounded gradient refinement (L-BFGS-B), the present framework removes this reliance on user-defined initial values and extends Gompertz-type modelling to multi-phase (polyauxic) profiles, while preserving numerical robustness.

## 6.2 Identification and exclusion of outliers

Reliable estimation of kinetic parameters requires not only robust model fitting but also rigorous assessment of data quality. Outliers are data points that substantially deviate from the prevailing data trend, which can bias parameter estimates, reduce predictive accuracy, and compromise biological interpretation. The presence of outliers is particularly problematic in nonlinear regression, as even a single aberrant value may unduly influence the fitted model.

When fitting kinetic models to experimental data, the presence of outliers can critically compromise the reliability of parameter estimation. The commonly used residual sum of squares (RSS) loss function assumes that measurement errors are normally distributed; however, biological datasets frequently contain outliers due to experimental variability or measurement artifacts. To address this, a two-stage fitting strategy is implemented. First, a robust loss function (Charbonnier et al. 1994; Barron 2019) is minimized during a pre-fit stage. This function reduces the influence of extreme deviations by penalizing large residuals linearly rather than quadratically.

Based on this robust pre-fit, outliers are formally identified using the Robust Regression and Outlier Removal (ROUT) method (Motulsky and Brown 2006), which utilizes the False Discovery Rate (FDR) to detect data points that are statistically inconsistent with the underlying model (Benjamini and Hochberg 1995). Upon the exclusion of these identified anomalies, the final parameter estimation is performed by minimizing the RSS on the cleaned dataset. This hybrid approach enhances the resilience of the model fitting procedure to outlier effects while preserving the statistical properties of the RSS for the final, biologically meaningful interpretation of the experimental data. The method moves beyond subjective or ad hoc exclusion, providing a statistically controlled, reproducible, and objective mechanism for detecting data points inconsistent with the majority distribution.

The theoretical foundation of the ROUT method aligns directly with the rationale for adopting the Charbonnier loss function for primary model fitting. Both approaches are based on the premise that, in biological datasets, most data points follow an underlying model, but deviations are more realistically described by a heavy-tailed (Charbonnier) distribution rather than a normal (Gaussian) distribution. Using a Charbonnier loss function inherently reduces the influence of outliers on parameter estimation, making the fit more robust. Building on this, the ROUT procedure formally identifies candidate outliers by quantifying the probability that each data point's deviation from the fitted model could result from random scatter, using the FDR framework. This allows explicit control over the proportion of false positives among flagged outliers: stricter FDR settings provide conservative exclusion, while less stringent thresholds increase sensitivity to detect atypical points.

The ROUT method thereby ensures that only data points statistically inconsistent with the modeled process are excluded, preserving the integrity of the main data structure while protecting model estimation from distortion. All excluded points are transparently reported, and the potential experimental or instrumental causes of outlier occurrence are addressed in the interpretation.

### 6.3 Adjusting number of stacked sigmoid

To prevent overparameterization and ensure model parsimony when fitting composite sigmoidal models to experimental data, the optimal number of growth patterns ($n$) should be determined using an established information criterion. For each candidate value of $n$, the model is fitted to the data, and the information criterion is calculated. The optimal $n$ is the smallest value for which the chosen criterion reaches its minimum; if the criterion remains constant or begins to increase with higher $n$, the lowest corresponding $n$ should be selected. Information criteria balance model quality by weighing goodness of fit against complexity, penalizing unnecessary parameters that do not substantially improve the model. This study considered three different information criteria: Akaike (Akaike 1974), Correlated Akaike (Hurvich and Tsai 1989) and Bayesian Information Criteria (Schwarz 1978). Table 1 lists the equations for the information criteria used in this study and defines the conditions for their application (Burnham and Anderson 2004).

Table 1 – Information criteria used for model selection and their respective formulations for assessing the trade-off between goodness-of-fit and complexity.

| Criteria | Canonical Form | | Condition |
|---|---|---|---|
| Akaike Information Criteria – AIC | $2 \cdot k - 2 \cdot \ln(\hat{L})$ | Equation 34 | $\frac{N}{k} \geq 40, N \leq 200$ |
| Correlated AIC – AICc | $AIC + \frac{2 \cdot k \cdot (k+1)}{(N-k-1)}$ | Equation 35 | $\frac{N}{k} < 40, N \leq 200$ |
| Bayesian Information Criteria – BIC | $k \cdot \ln(N) - 2 \cdot \ln(\hat{L})$ | Equation 36 | $\frac{N}{k} \gg 40, N > 200$ |

where $k$ is the total number of fitted parameters (cardinality of vector $\theta$); $N$ is the number of the points in experimental data, and $\hat{L}$ denotes the maximum likelihood of the fitted model. $\hat{L}$ is the highest value of the likelihood function evaluated at the parameter estimates that best fit the data, and in the context of nonlinear regression with independent, normally distributed errors of constant variance, as shown in Equation 37.

$$\hat{L}(\hat{\theta}) = \prod_{m=1}^{N} \frac{1}{\sqrt{2 \cdot \pi \cdot \sigma^2}} \cdot e^{-\frac{\left(y_m - \hat{y}(x_m;\hat{\theta})\right)^2}{2 \cdot \sigma^2}} \qquad \text{Equation 37}$$

where $\sigma$ is the standard deviation of the residual noise. The standard Akaike Information Criterion (AIC) is considered reliable for balancing model fit and complexity. The use of the corrected AIC (AICc) is strongly recommended to compensate for small-sample bias and to avoid favoring overparameterized models. The Bayesian Information Criterion (BIC), which imposes a stronger penalty for model complexity as the sample size increases, can be applied for any sample size but is especially appropriate when prioritizing model parsimony in large datasets. In practice, AIC and AICc are more sensitive to predictive accuracy, while BIC tends to select simpler models as $N$ increases. Therefore, for large $N$ and a strong emphasis on parsimony, BIC may be favored.

### 6.4 Estimating parameters errors

After successful model fitting, it is essential to quantify the uncertainty of each estimated parameter. The standard errors of the fitted parameters can be approximated using the Hessian matrix of the loss function evaluated at the optimum parameter vector, $\hat{\theta}$. While the Jacobian matrix contains first derivatives and describes the local slopes of the loss surface, the Hessian consists of second-order partial derivatives, providing information on the local curvature around the minimum of the loss function. Thus, Hessian matrix yields a more accurate estimate of parameter uncertainty, particularly for highly nonlinear models. However, as analytical derivatives are complex, the Hessian is estimated numerically using central finite differences.

The Hessian matrix of the loss function with respect to optimum parameter vector $\hat{\theta}$ is given by Equation 38:

$$H_{ij} = \frac{\partial^2 RSS(\hat{\theta})}{\partial \hat{\theta}_i \partial \hat{\theta}_j} \quad \text{Equation 38}$$

When the Hessian matrix $H_{ij}$ is ill-conditioned or singular, the standard inverse $H_{ij}^{-1}$ is undefined. To ensure numerical stability and obtain a valid covariance matrix $C_{ij}$, is employed the Moore-Penrose pseudo-inverse ($H_{ij}^{+}$), computed through Singular Value Decomposition (Penrose 1955). At the optimum parameter vector ($\hat{\theta}$), the covariance matrix $C$ is defined as the pseudo-inverse of the Hessian scaled by the estimated residual variance ($\sigma^2$):

$$C_{ij} = \sigma^2 \cdot H_{ij}^{+} \quad \text{Equation 39}$$

The term $\sigma^2$ represents the approximate estimated residual variance, which can be approximated as shown in Equation 40:

$$\sigma^2 \approx \hat{\sigma}^2 = \frac{RSS(\hat{\theta})}{N - k} \quad \text{Equation 40}$$

Finally, the standard error of parameter $\theta_j$ is the square root of the corresponding diagonal element of the covariance matrix.

$$SE_{\theta_j} = \sqrt{C_{jj}} \quad \text{Equation 41}$$

For derived parameters, such as the phase fractions ($p_j$) calculated via the Softmax function, standard errors are approximated using the Delta method (Oehlert 1992). This involves propagating the covariance of the latent variables ($z$) through the Jacobian of the Softmax transformation, ensuring that uncertainty is correctly quantified for all biologically relevant parameters. This approach provides a principled estimate of the uncertainty, supporting robust statistical inference and reliable interpretation of the fitted results.

### 6.5 Model Applicability

While structured and cybernetic models (Kompala et al. 1986; Ramkrishna and Song 2012; Bate et al. 2023) provide a detailed description of intracellular regulatory mechanisms, their practical application in environmental biotechnology and industrial bioprocesses is often constrained by the limited observability of internal state variables. In experimental setups involving complex substrates and mixed cultures, data are frequently restricted to macroscopic or indirect measurements (e.g.: cumulative product formation, total substrate consumption, biomass concentrations expressed as volatile solids concentration) without access to intracellular enzyme concentrations or specific mRNA levels required to calibrate complex metabolic models. Consequently, the proposed semi-mechanistic framework fills a gap by offering a robust engineering tool that extracts physiologically interpretable kinetic parameters (e.g., maximum specific rates and phase durations) directly from standard input-output data, avoiding the over-parameterization and experimental prohibitiveness associated with fully structured models in macroscopic process optimization.

## 7 Computational Implementation and Software Availability

To ensure full reproducibility, transparency, and broad community adoption, the complete computational workflow described in this study, including the polyauxic regression models, the hybrid optimization pipeline (Differential Evolution coupled with L-BFGS-B), and the outlier detection algorithms, has been implemented as a modular open-source Python package (Mockaitis 2025), as summarized in Figure 6. The repository contains the source code, comprehensive documentation, and the specific scripts required to reproduce all figures and analyses presented in this manuscript.

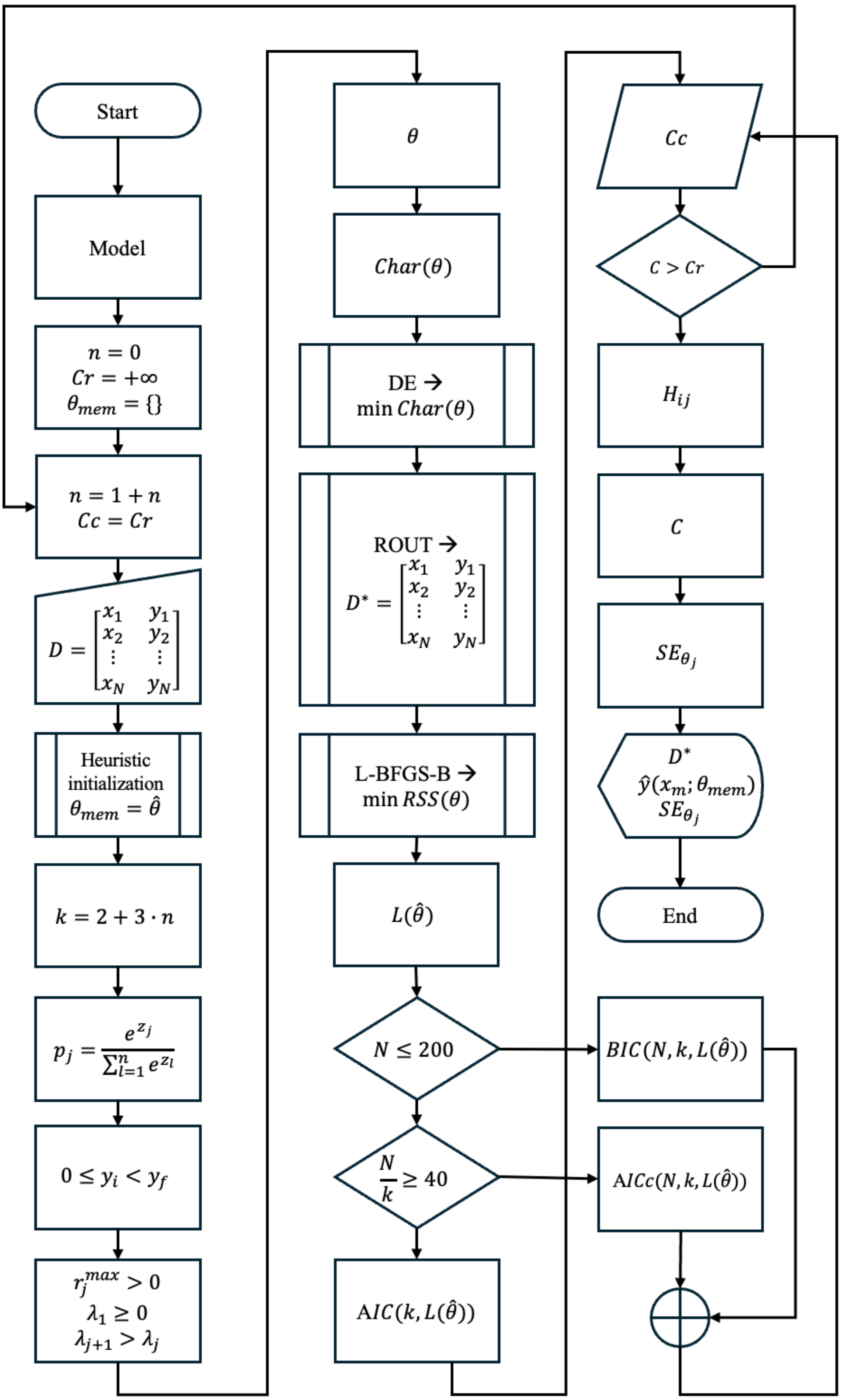

Figure 6 – Computational flowchart of the proposed semi-mechanistic framework. The pipeline integrates heuristic initialization, robust global optimization using Differential Evolution (DE), outlier detection via the ROUT method, and local refinement using L-BFGS-B.

The algorithmic variables presented in Figure 6 are defined as follows: $D$ represents the raw experimental dataset with sample size $N$; $D^*$ denotes the cleaned dataset after outlier removal via the ROUT method. The variables $x_m$ e $y_m$ correspond to the independent (time) and dependent (response) variables, respectively. The counter $n$ indicates the number of auxic growth phases, determining the total number of parameters $k$ (where $k = 2 + 3 \cdot n$). The parameter vector is denoted by $\theta$, with $\hat{\theta}$ representing the estimated values. The weighting factors $p_j$ are derived from unconstrained latent variables $z_j$ via a Softmax transformation. $Char(\theta)$ is the Charbonnier robust loss function used in the global Differential Evolution (DE) search; $RSS(\theta)$ is the Residual Sum of Squares minimized during the L-BFGS-B local refinement. $L(\hat{\theta})$ is the maximum likelihood of the fitted model. AIC, AICc, and BIC are the Akaike, Corrected Akaike, and Bayesian Information Criteria, respectively. The variables $Cr$ (Reference Criterion) and $Cc$ (Current Criterion) track the optimization progress, while $\theta_{mem}$ stores the parameter set of the best parsimonious model found to date. $H_{ij}$ is the Hessian matrix used to compute the covariance matrix $C$ and the standard errors $SE_{\theta_j}$ for each parameter.

For archival stability and traceability, the software version corresponding to the results reported herein is preserved as Release v1.0.0. The code and datasets can be accessed at the following repository:

- Source Code & Documentation: https://github.com/gusmock/mono_polyauxic_kinetics/
- v1.0.0 Release https://github.com/gusmock/mono_polyauxic_kinetics/releases/tag/v1.0.0

Additionally, to facilitate exploratory use without the need for local installation or programming expertise, an interactive web-based demonstration of the platform is available at:

- Web Application: https://monopolyauxickinetics-test.streamlit.app

There are existing tutorials for Gompertz fitting which are typically tied to specific proprietary platforms, focusing on single-phase, non-overlapping growth curves (Guo and Wang 2024). The present implementation encapsulates the entire polyauxic workflow in an open, platform-agnostic Python package and an interactive web application. This design facilitates reproducible analyses of complex multiphasic datasets and lowers the barrier for applying advanced optimisation strategies (DE + L-BFGS-B, ROUT outlier detection, multi-criteria phase selection) in routine bioprocess studies.

## 8 Results and Discussion

The polyauxic framework was applied to experimental datasets from the literature assessing the kinetics of methane production from aquatic macrophytes under hydrothermal pretreatment (Ferraz Dutra et al. 2024). The original study employed a first-order kinetic model, a common approach in anaerobic digestion that serves as an approximation of Monod kinetics under limiting substrate concentrations (Mockaitis et al. 2006, 2012). To facilitate a direct comparison of kinetic parameters, the experimental data were re-fitted using the monoauxic Boltzmann and Gompertz models (Table 2). Additionally, the first-order model was re-parameterized to align with the nomenclature of the Boltzmann and Gompertz equations (Equation 42). Since the standard first-order model does not account for lag phases or polyauxic growth, the comparison focused primarily on the maximum methane production ($y_f$) and the maximum reaction rate ($r_1^{max}$).

$$y(x) = y_f \cdot \left(1 - e^{-r_1^{max} \cdot t}\right) \qquad \text{Equation 42}$$

The first-order model employed in the study assumes no initial methane production ($y_i = 0$), whereas the framework typically estimates $y_i$ as a free parameter. However, a dual fitting strategy was adopted for comparison, with $y_i$ allowed to float, and with $y_i$ constrained to zero. Table 2 presents the kinetic parameters through Equation 42 (Ferraz Dutra et al. 2024), and using the monoauxic ($n = 1$) Boltzmann (Equation 31) and Gompertz (Equation 32) models.

Table 2 – Comparison of kinetic parameters for methane production from aquatic macrophytes derived from first-order, Boltzmann, and Gompertz models. Values in parenthesis denote fits where the initial methane production ($y_i$) was fixed at zero.

| Macrophyte | Parameter | 1st order | Monoauxic Boltzmann | Monoauxic Gompertz |
|---|---|---|---|---|
| *E. densa* | $y_i$ | 0 | 7.7 ± 17.8 (0) | 10.1 ± 13.7 (0) |
| | $y_f$ | 154.6 | 144.5 ± 2.8 (144.2 ± 2.8) | 145.5 ± 2.6 (144.8 ± 2.7) |
| | $r_1^{max}$ | 26 | 24 ± 3 (27 ± 4) | 25 ± 3 (29 ± 4) |
| | $R^2$ | 0.96 | 0.9266 (0.9242) | 0.9416 (0.9375) |
| *E. crassipes* | $y_i$ | 0 | 5.2 ± 9.8 (0) | 7.8 ± 8.9 (0) |
| | $y_f$ | 133.0 | 108.7 ± 2.6 (108.1 ± 2.6) | 111.8 ± 3.1 (110.0 ± 2.9) |
| | $r_1^{max}$ | 10 | 9.1 ± 0.7 (10 ± 1) | 9.2 ± 0.7 (10 ± 1) |
| | $R^2$ | 0.97 | 0.9653 (0.9621) | 0.9682 (0.9612) |
| *T. domingensis* | $y_i$ | 0 | 49.8 ± 3.0 (0) | 18.1 ± 14.0 (0) |
| | $y_f$ | 140.7 | 113.7 ± 2.3 (118.0 ± 4.0) | 123.0 ± 5.1 (118.8 ± 4.6) |
| | $\lambda_1$ | - | 6.5 ± 0.8 (-) | - |
| | $r_1^{max}$ | 11 | 22 ± 11 (11 ± 1) | 8.8 ± 1.0 (12 ± 1) |
| | $R^2$ | 0.93 | 0.9393 (0.8939) | 0.9221 (0.8859) |
| *C. papyrus nanus* | $y_i$ | 0 | 13.1 ± 5.4 (0) | 19.1 ± 4.0 (0) |
| | $y_f$ | 110.5 | 92.0 ± 1.7 (93.4 ± 4.8) | 92.3 ± 2.1 (96.0 ± 1.1) |
| | $\lambda_1$ | - | 2.6 ± 1.0 (0.5 ± 6.8) | 3.4 ± 0.8 (0.7 ± 2.2) |
| | $r_1^{max}$ | 13 | 13 ± 2 (11 ± 4) | 14 ± 3 (10 ± 1) |
| | $R^2$ | 0.93 | 0.9638 (0.9603) | 0.9585 (0.9512) |

The application of the proposed monoauxic Boltzmann and Gompertz models to literature data allows for a direct comparison with the standard first-order kinetic model commonly used in anaerobic digestion. As shown in Table 2, the first-order model inherently assumes zero initial production ($y_i = 0$). Consequently, when the monoauxic models are constrained to this same assumption (values in parentheses), the estimated maximum yield ($y_f$) and maximum rates ($r_1^{max}$) are statistically similar across all models for substrates with simple degradation profiles, such as *E. densa* and *E. crassipes*. However, significant discrepancies arise for substrates exhibiting more complex kinetic behaviors, specifically *T. domingensis* and *C. papyrus nanus*. For *T. domingensis*, the unconstrained monoauxic fit resulted in an unrealistically $y_i$ value, whereas forcing it to zero caused a notable drop in the coefficient of determination. This sensitivity indicates that forcing the intercept to zero is valid only when the lag phase is negligible (pseudo-first-order behavior) in monoauxic patterns. As methane production for *T. domingensis* and *C. papyrus nanus* is likely polyauxic, trying to fit an inappropriate monoauxic model can result in recognizing valid experimental data as outliers and determining unrealistic kinetic parameters. Therefore, when a genuine lag phase exists or when growth shows a polyauxic behavior, forcing $y_i = 0$ can cause the optimization algorithm to compensate by estimating a mathematically negative (and biologically meaningless) lag phase to fit the curve's slope.

Furthermore, the kinetic profiles of *T. domingensis* and *C. papyrus nanus* highlight the structural limitations of first-order and simple monoauxic kinetics. While the monoauxic Boltzmann and Gompertz models estimated negligible lag phases ($\lambda_1 \approx 0$) for *E. densa* and *E. crassipes*, indicating immediate methane production, significant lag phases times (ranging from 8.8 to 14 h) were detected for *T. domingensis* and *C. papyrus nanus*. This observation suggests the presence of either a genuine underlying lag phase or a polyauxic methane production pattern, phenomena the standard first-order model is structurally unable to capture. More importantly, visual heuristic analysis of the experimental data for these two macrophytes suggests the presence of multiple growth patterns. The inability of the first-order model to account for a lag phase, combined with the difficulty of the monoauxic models to fit the data without floating the intercept, strongly suggests that these substrates undergo polyauxic degradation. In such cases, the observed lag time may partially represent the transition between different metabolic phases that neither first-order nor simple monoauxic models can fully resolve.

To mitigate the subjectivity inherent in heuristic visual analysis, the computational pipeline was configured to evaluate kinetic models ranging from monoauxic to 10-auxic phases. However, as the experimental datasets lacked replicates and contained limited observations ($N \approx 20$), the algorithm could not converge for models exceeding 5 phases due to insufficient degrees of freedom (overparameterization). Data from all experiments also shows that no lag phase was observed at the beginning of each assay, thus $y_i$ were constrained to zero in all fittings performed. As the cardinality of parameter vector is $k = (3n + 2)$ for both Boltzmann and Gompertz models (where $n$ is the number of growth phases) the ratio $N/k = [3.8 - 1.1]$, AICc is the criterion of choice for all tested models. Table 3 shows the behaviour of AICc for the tested kinetic models for all studied aquatic macrophytes.

Table 3 – Comparison of AICc values obtained for monoauxic and polyauxic kinetic models applied to methane production data. The optimal model corresponds to the first local minimum of the AICc function (in red).

| n | $\frac{N}{k}$ | Correlated Akaike Information Criteria – AICc | | | | | | | |
|---|---|---|---|---|---|---|---|---|---|
| | | *E. densa* | | *E. crassipes* | | *T. domingensis* | | *C. papyrus nanus* | |
| | | Boltzmann | Gompertz | Boltzmann | Gompertz | Boltzmann | Gompertz | Boltzmann | Gompertz |
| 1 | 3.8 | 106.3 | 102.7 | 85.8 | 86.2 | 106.0 | 107.4 | 86.2 | 90.3 |
| 2 | 2.4 | 112.7 | 110.8 | 76.6 | 91.9 | 95.0 | 94.0 | 95.9 | 85.7 |
| 3 | 1.7 | 124.4 | 140.2 | 122.0 | 122.1 | 109.4 | 106.7 | 120.5 | 118.4 |
| 4 | 1.4 | 197.0 | 197.4 | 195.2 | 194.2 | 178.3 | 180.0 | 176.2 | 175.0 |
| 5 | 1.1 | 712.3 | 713.2 | 708.2 | 707.9 | 690.9 | 691.2 | 401.5 | 391.3 |

Table 3 indicates that for *E. densa* and *T. domingensis*, the information criteria for both Boltzmann and Gompertz models converge, suggesting monoauxic and diauxic behaviors, respectively. However, for *E. crassipes* and *C. papyrus nanus*, the criteria diverge, leading to different phase estimations depending on the model structure. Figure 7 shows these optimum fitting results for Boltzmann and Gompertz models for each studied macrophytes.

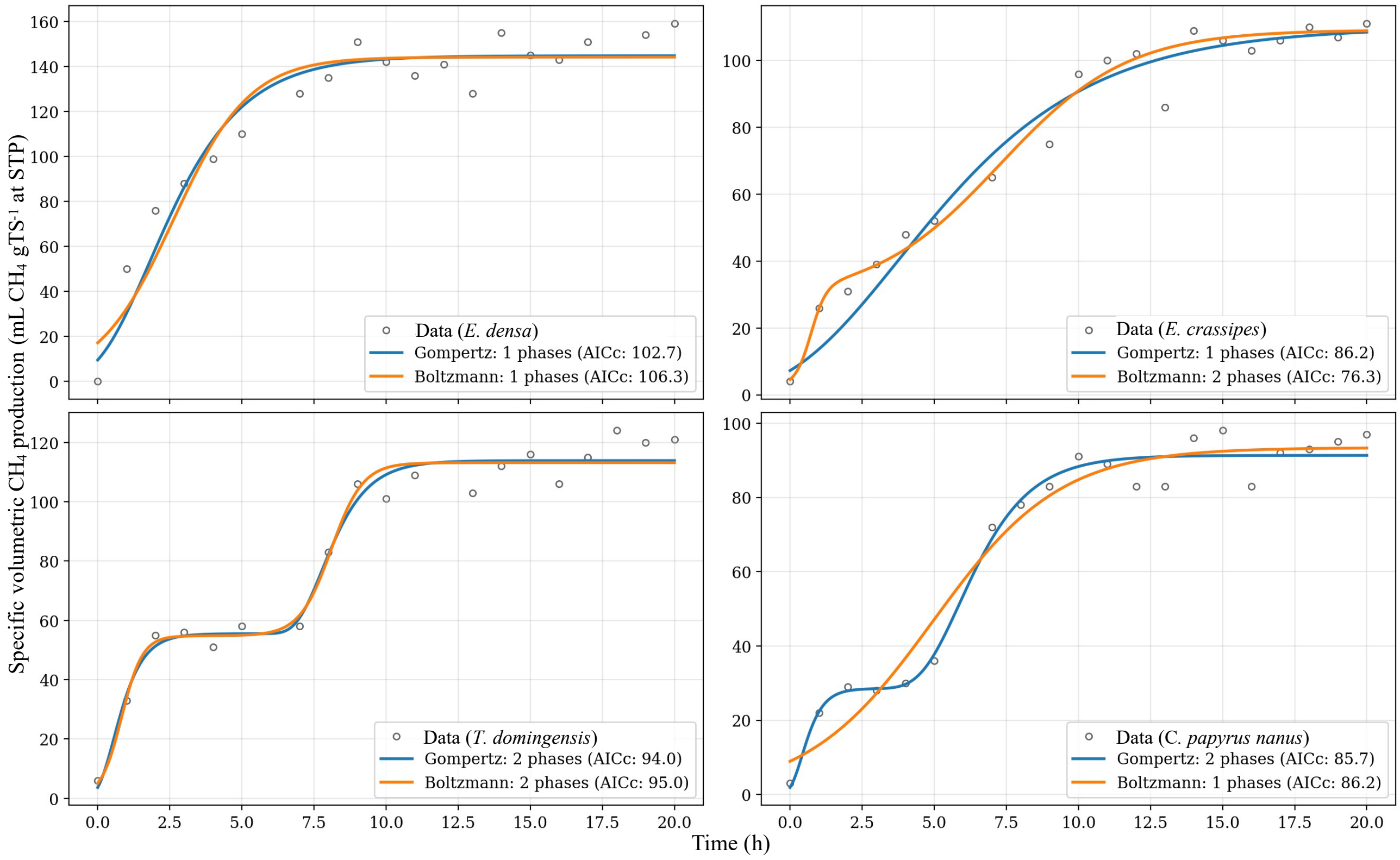

Figure 7 – Experimental data for specific volumetric methane production (relative to applied total solids – TS) and optimal model fits (Boltzmann or Gompertz) based on the AICc selection criteria presented in Table 3.

The discrepancy between both models when fitting *E. crassipes* and *C. papyrus nanus* methane production data stems from a fundamental structural difference. Boltzmann model is strictly symmetrical around its inflection point, whereas the Gompertz model is inherently asymmetrical. Because standard first-order kinetics describe an immediate, decelerating process ($x = 0$, no lag phase) which is highly asymmetrical, the symmetrical Boltzmann equation often forces a poor fit or yields unrealistic parameters. In contrast, the asymmetry of the Gompertz model allows it to approximate pseudo-first-order dynamics effectively. Therefore, when criteria disagree or when substrates exhibit negligible lag phases, the polyauxic Gompertz model (particularly with a constrained intercept $y_i = 0$) is the superior mechanistic choice. This mechanistic preference holds even when information criteria strictly favor an alternative, as observed with *E. crassipes*, where AICc indicated a preference for a diauxic Boltzmann model over a monoauxic Gompertz. In such cases, the statistical preference for the more complex Boltzmann model is likely an artifact, where the algorithm introduces an artificial second phase solely to compensate for the model's inability to fit the asymmetrical, first-order-like data with a single symmetrical curve (Mangan et al. 2017).

In conclusion, while mechanistic plausibility favors the Gompertz model for asymmetrical, first-order-like kinetics, the robustness of these findings is fundamentally limited by the absence of experimental replicates. For instance, the monoauxic Boltzmann model successfully fitted methane production for *C. papyrus nanus*, but only by introducing a large standard error in the initial methane production parameter ($y_i = 0.0 \pm 44.9$). Although the omission of replicates is often justified when designing assays involving complex mixed cultures (such as biomethanogenic assays) due to the inherent stochasticity and unique community dynamics of each experimental unit (Vaux et al. 2012; Goux et al. 2015), this constraint poses a specific challenge for kinetic modeling. Without the statistical averaging provided by replicates, distinguishing between genuine physiological phenomena (such as polyauxic shifts) and experimental noise becomes difficult. Consequently, outlier exclusion becomes a sensitive issue, increasing the risk of overfitting complex models to singular stochastic events rather than robust biological trends.

The effect of substrate composition on kinetic parameters using a modified Boltzmann function for diauxic growth was previously investigated (Volpi et al. 2022). While the use of biological triplicates allowed for the assessment of assay variability, the study highlighted significant modeling challenges. Notably, the initial growth phase for several substrates exhibited pseudo-first-order behavior with negligible lag phases, exposing the structural limitation of the Boltzmann model, which strictly imposes a symmetrical sigmoidal curve around an inflection point and consequently struggles to fit asymmetrical, front-loaded data. In this context, kinetic parameters were compared for assays using poultry slaughterhouse sludge as inoculum, specifically for the mono-digestion of vinasse and the co-digestion of vinasse, deacetylation liquor, and filter cake. Table 4 presents a comparison between the parameters reported by the study authors and the values obtained in this work.

Table 4 – Comparison of kinetic parameters reported using a diauxic Boltzmann model (Volpi et al. 2022) versus those obtained in this study using an n-auxic model for single-batch anaerobic digestion.

| Parameter | | Vinasse (theoretical $y_f = 548$) | | | Codigestion (theoretical $y_f = 718$) | | |
|---|---|---|---|---|---|---|---|
| | | **Diauxic Boltzmann** | **n-auxic Boltzmann** | **n-auxic Gompertz** | **Diauxic Boltzmann** | **n-auxic Boltzmann** | **n-auxic Gompertz** |
| $n$ | | 2 | 4 | 4 | 2 | 2 | 2 |
| $y_f$ | | - | 0 ± 11 | 0 ± 17 | - | 0 ± 37 | 0 ± 86 |
| $y_i$ | | 513 ± 3 | 508 ± 2 | 510 ± 2 | 717 ± 119 | 731 ± 109 | 900 ± 247 |
| $p$ | $p_1$ | 0.44 ± 0.01 | 0.20 ± 0.04 | 0.22 ± 0.07 | 0.32 ± 0.02 | 0.26 ± 0.09 | 0.25 ± 0.13 |
| | $p_2$ | 0.56 | 0.27 ± 0.04 | 0.29 ± 0.07 | 0.68 | 0.74 ± 0.09 | 0.75 ± 0.13 |
| | $p_3$ | - | 0.31 ± 0.05 | 0.28 ± 0.03 | - | - | - |
| | $p_4$ | - | 0.22 ± 0.04 | 0.21 ± 0.02 | - | - | - |
| $\lambda$ | $\lambda_1$ | 11 ± 1 | 0.70 ± 1.07 | 0.03 ± 1.46 | 13 ± 2 | 0 ± 6 | 0 ± 11 |
| | $\lambda_2$ | 75 ± 0 | 3.72 ± 4.45 | 3.66 ± 7.59 | 94 ± 8 | 60 ± 5 | 63 ± 4 |
| | $\lambda_3$ | - | 60.2 ± 1.2 | 61.8 ± 0.9 | - | - | - |
| | $\lambda_4$ | - | 77.7 ± 1.0 | 77.7 ± 0.5 | - | - | - |
| $r^{max}$ | $r_1^{max}$ | 7.5 ± 0.4 | 12.3 ± 1.2 | 12.5 ± 1.6 | 7.2 ± 1.6 | 7.63 ± 1.37 | 8.19 ± 1.50 |
| | $r_2^{max}$ | 11 ± 1 | 3.88 ± 0.38 | 3.43 ± 0.56 | 7.7 ± 0.7 | 7.64 ± 0.52 | 7.39 ± 0.42 |
| | $r_3^{max}$ | - | 8.40 ± 0.76 | 8.53 ± 0.63 | - | - | - |
| | $r_4^{max}$ | - | 14.9 ± 1.8 | 16.4 ± 2.0 | - | - | - |
| $R^2$ | | 0.99 | 0.9942 | 0.9941 | 0.90 | 0.9014 | 0.9014 |

The reliability of the kinetic parameters presented in Table 4 is significantly reinforced by the experimental strategy employed. A comparative analysis indicates that the use of biological replicates, as performed by authors (Volpi et al. 2022), contributes decisively to the stability and reliability of the estimated parameters. This stands in contrast to the methane production potential from aquatic macrophytes, which relied on single-replicate assays; the inclusion of replicates in the present work minimizes experimental artifacts and ensures that the derived kinetics reflect reproducible biological trends rather than stochastic variations (Ferraz Dutra et al. 2024).

Furthermore, the application of the n-auxic model revealed distinct behaviors governed by substrate complexity. In the case of co-digestion, the increased heterogeneity of the substrate matrix likely resulted in overlapping metabolic pathways. Despite this complexity, the n-auxic model successfully described the process by aggregating these concomitant phases, effectively fitting the data without unnecessarily increasing the model's structural complexity. Conversely, while vinasse retains its own biochemical intricacies, it presents a simpler matrix compared to the co-digestion mixtures. This relative simplicity facilitated the identification of clearer, more distinct metabolic phases, allowing for a more highly parametrized model that could resolve specific shifts in methane production that are otherwise obscured in complex co-digestion scenarios.

Hidden growth patterns could exist within the reported kinetic profiles, representing a distinct metabolic transition masked by the limitations of the double sigmoidal model applied in the study. While the authors attributed the observed diauxic behavior to secondary microbial species growing on metabolic products, the mass balance analysis revealed that predicted methane production values often exceeded experimental measurements, indicating the presence of active but unconsidered metabolic routes. In such a complex consortium, where intermediates like ethanol, propionate, and isovalerate are produced and subsequently consumed, a specialized sub-population scavenging a specific transient intermediate could generate more patterns of microbial activity. However, the fitting was restricted to a superposition of two Boltzmann functions, this intermediate phase would likely be absorbed into the broader lag or stationary regions of the dominant phases, rendering it statistically invisible without the application of a polyauxic model to resolve the overlapping kinetics.

To illustrate the robustness of this multiphasic resolution, Figure 8 compares the 4 phase polyauxic Gompertz and Boltzmann models in fitting the vinasse experimental data. The figure details the exclusion of detected outliers and displays the mean production values with standard deviations, alongside the decomposition of the cumulative curve into its individual sigmoidal phases.

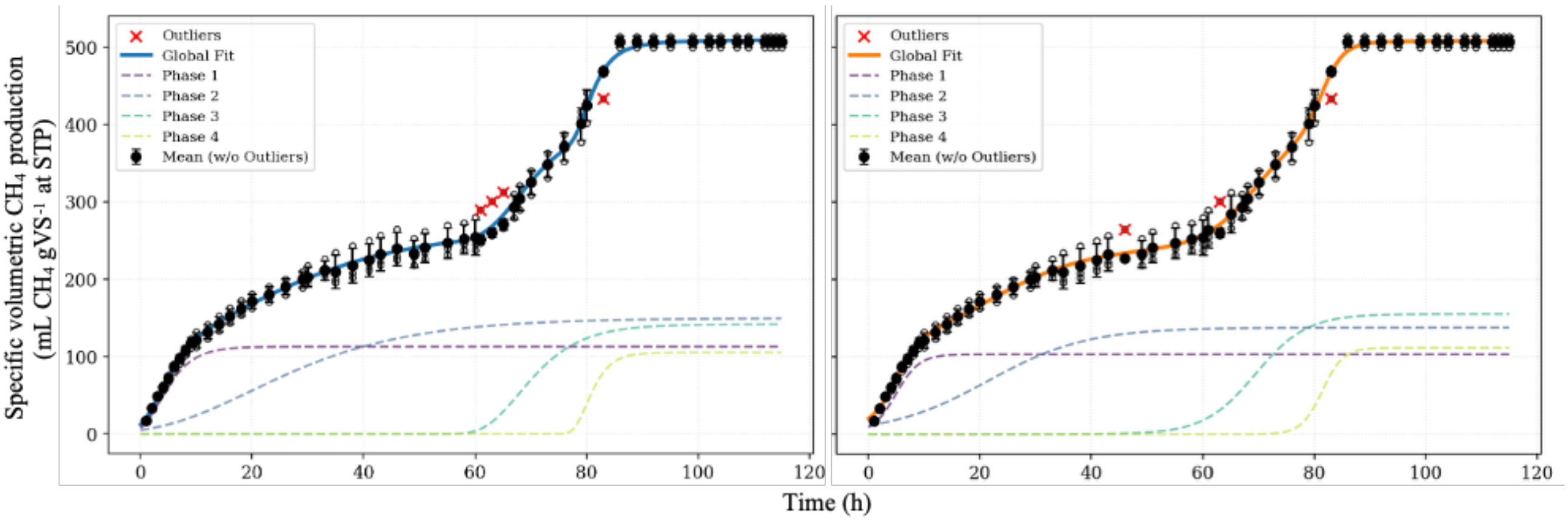

Figure 8 – Experimental data for specific volumetric methane production (relative to applied volatile solids – VS) using vinasse as substrate and the optimal model fitting for polyauxic Gompertz (in blue line) and Boltzmann (in orange line). Data points marked with a red 'X' represent outliers.

The proposed kinetic framework was also applied to the more complex co-digestion assay (mixture of vinasse, deacetylation liquor, and filter cake) to assess model performance in a highly heterogeneous matrix, as depicted in Figure 9.

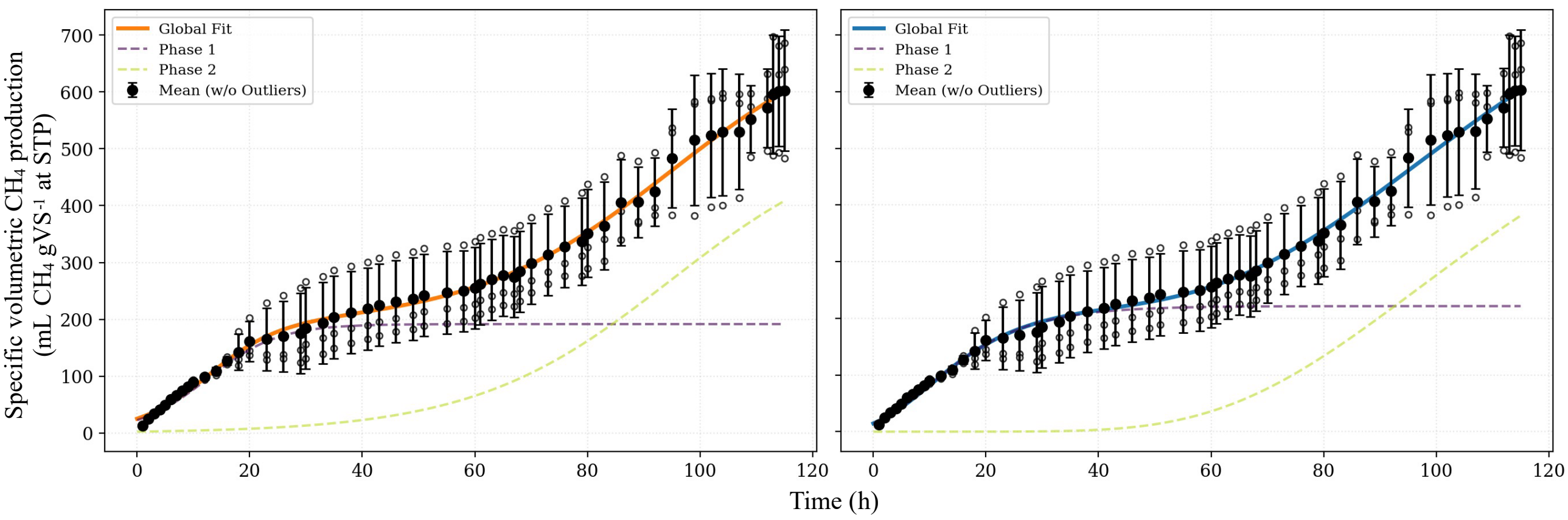

Figure 9 – Experimental data for specific volumetric methane production (relative to applied volatile solids – VS) from vinasse, deacetylation liquor and filter pie codigestion and the optimal model fitting for polyauxic Gompertz (in blue line) and Boltzmann (in orange line).

A critical limitation arising from the experimental protocol is highlighted in Figure 9. Unlike the experiment using vinasse as sole substrate (Figure 8), the co-digestion assay failed to reach a distinct stationary phase (plateau). Although the assays were terminated strictly according to the VDI 4630 standard (when daily production dropped below 1% of the accumulated volume), the kinetic model predicted methane production continuing well beyond this timeframe. This discrepancy exposes a fundamental flaw in relying solely on VDI 4630 for kinetic modeling, since by enforcing an artificial endpoint, the protocol classifies the experiment as unfinished relative to the biological potential. Consequently, optimization algorithms are blinded to the true maximum yield or any other growth phases. To mitigate this, one must either mathematically fix the parameter to the final experimental value to force convergence or accept estimates plagued by substantial standard errors due to the absence of a defined asymptote. Thus, employing VDI 4630 without adaptation effectively reduces kinetic modeling from a predictive tool for biological behavior to a mere curve-fitting exercise constrained by procedural artifacts.

This lack of a defined plateau creates a significant divergence in model predictions. In the co-digestion assays, the Gompertz model predicted a significantly higher methane production potential than the Boltzmann model. While the Boltzmann prediction aligned more closely with the theoretical values, this alignment may be misleading. As noted by the authors, the deacetylation liquor experiment likely resulted in an underestimated $y_f$ value. Therefore, the Gompertz model's higher prediction might accurately reflect the unobserved potential, while the Boltzmann model merely mirrors the premature experimental termination. This ambiguity suggests that the abrupt termination of the experiment affects model representativity far more severely than the lack of replicates; without a plateau, it becomes mathematically impossible to objectively discriminate between models with such diverging predictions.

Furthermore, Figure 8 demonstrates that after 20 hours of digestion, the substrate complexity resulted in a substantial standard deviation between samples. Notably, the algorithm did not identify or exclude any outliers in this assay, in contrast to the vinasse experiment. While this high variability complicates the resolution of distinct underlying phases, it is methodologically superior to retain the full dataset rather than artificially reducing noise. Relying on the model’s robust criteria for phase definition and outlier detection ensures a more statistically representative fit compared to single-replicate designs, even if the resulting variance challenges the idealized representativeness often sought in literature.

Consequently, selecting between the Boltzmann and Gompertz models requires a multi-faceted evaluation. Beyond assessing whether the initial growth pattern exhibits a first-order-like kinetic profile (characterized by the absence of a lag phase and an initial maximum reaction rate), the decision must prioritize model parsimony to avoid over-parametrization. Furthermore, the selection must weigh the number of outliers excluded by the ROUT method (with the specified FDR) against the results of the chosen information criteria.

Ultimately, despite these challenges, the polyauxic modeling framework demonstrates clear superiority over stand-alone, single-phase fittings. By acknowledging the multiphasic nature of complex substrates, the n-auxic approach captures the aggregation of concomitant metabolic reactions that simple models invariably miss. It transforms the high variability of complex growth patterns into structured, interpretable phases, offering a mechanistic insight that single-equation fits cannot provide. However, it is important to acknowledge that this approach does not entirely eliminate subjectivity. The decision of how many phases to include, balancing mathematical precision with biological reality, remains a form of heuristic parsimony. Therefore, the interpretation of the most representative fit must consider not only statistical metrics but also biological plausibility, ensuring that the added complexity of the n-auxic model explains the data rather than merely overfit it.

## 9 Conclusion

This work establishes a rigorous mathematical framework for modeling monoauxic and polyauxic microbial growth by reformulating classic Boltzmann and Gompertz equations with biologically interpretable parameters. The proposed methodology integrates a robust computational workflow (combining Differential Evolution, Charbonnier loss minimization, and the ROUT outlier detection method) to overcome the challenges of parameter non-identifiability in high-dimensional spaces.

The application of this framework to anaerobic digestion as an example of high complexity growth kinetics datasets revealed fundamental limitations in standard experimental protocols. The reliance on artificial termination criteria, such as those in VDI 4630, was shown to mislead optimization algorithms, reducing kinetic modeling to a mere curve fitting. Furthermore, the analysis highlighted that while statistical criteria (AIC/BIC) prevent overparameterization, the distinction between genuine multiphasic growth and experimental noise in complex co-digestion matrices relies heavily on the availability of biological replicates.

This study demonstrates that n-auxic modeling offers a semi-mechanistic window into complex substrate degradation that single-phase models miss. However, the selection of the most representative model remains a form of heuristic parsimony, requiring researchers to balance statistical precision with biological plausibility. The proposed workflow thus serves not just as a fitting tool, but as a comprehensive strategy for interpretable bioprocess analysis. Ultimately, this approach bridges the gap between empirical curve fitting and mechanistic Monod-type kinetics, providing a reproducible route for determining kinetic parameters, thereby serving as a valuable tool for environmental biotechnology and biorefinery operations.

## 10 Generative AI use statement

Generative AI (ChatGPT, OpenAI) was used exclusively to assist with text structuring and to review grammar and spelling. All scientific content, data interpretation, and conclusions are entirely the author's own.

## 11 Acknowledgements

The author gratefully acknowledges his mentors, especially Prof. Walter Borzani (*in memoriam*), Eugenio Foresti, Marcelo Zaiat, José Alberto Domingues Rodrigues, and Serge Roger Guiot, for their invaluable guidance to himself and generations of scientists in the study of microbial growth kinetics in complex systems.

## 12 Disclosure statement

The author declares no potential conflicts of interest.

## 13 Funding

This work was supported by the São Paulo Research Foundation (FAPESP), grant number 2018/18802-0, and by the National Council for Scientific and Technological Development (CNPq), grant numbers 425241/2018-1 and 409407/2023-2.